\documentclass[authoryear,preprint,a4paper,12pt]{elsarticle}

\usepackage{amssymb}
\usepackage{amsmath}
\usepackage{gensymb}
\usepackage[usenames,dvipsnames]{xcolor}
\usepackage{hyperref}
\usepackage{array}
\usepackage{multirow}
\usepackage{tabularx}
\usepackage{subfigure}
\usepackage{pdflscape}
\usepackage{setspace}
\usepackage{multirow}  
\usepackage{lineno}
\usepackage[top=2cm, bottom=2cm, left=2cm, right=2cm]{geometry}  
\usepackage{sectsty}

\newcommand   {\ds}[1]          {\displaystyle{#1}}
\newcommand   {\calV}           {\mathcal V}

\newcommand   {\calA}           {\mathcal A}

\newcommand   {\diff}           {{\rm\,\,d}}
\newcommand   {\dert}[1]        {\frac{\partial #1}{\partial t}}

\newcommand   {\q}              {\mbox{\boldmath $q$}}

\newcommand   {\ap}[1]          {\ifmmode^{\rm #1}\else$^{\rm #1}$\fi}
\usepackage{color}
\usepackage[normalem]{ulem}

\def\dout{\bgroup
 \markoverwith{\lower-0.2ex\hbox
 {\kern-.03em\vbox{\hrule width.2em\kern0.45ex\hrule}\kern-.03em}}%
 \ULon}
\MakeRobust\dout

\newcount\ndots \def\drwln#1#2{\raise 2.5pt\vbox{\hrule width #1pt
    height #2pt}}

 \def\square {${\vcenter{\hrule height
      .4pt \hbox{\vrule width .4pt height 3pt \kern 3pt \vrule width
        .4pt} \hrule height .4pt}}$\nobreak\ }  
   \def\filsqr {${\vcenter{\hrule height 2pt
      \hbox{\vrule width 2.2pt height 0.2pt \kern 0.1pt \vrule width
        2.2pt} \hrule height 2.2pt}}$\nobreak\ }

\definecolor{forest_green}{RGB}{34,139,34}
\definecolor{cyan}{RGB}{0,255,255}

\newcolumntype{C}[1]{>{\centering\let\newline\\\arraybackslash\hspace{0pt}}m{#1}}

\journal{Aerospace Science and Technology}

\graphicspath{{Figures/}}

\begin{document}
\allsectionsfont{\sffamily}

\begin{frontmatter}

\title{{\sffamily{\LARGE \textbf{Hydrodynamic analysis of the water landing phase of aircraft fuselages at constant speed and fixed attitude}}}}


\author[INM]{Emanuele Spinosa\corref{cor1}}
\ead{emanuele.spinosa@inm.cnr.it}

\author[INM]{Riccardo Broglia}

\author[INM]{Alessandro Iafrati}

\address[INM]{CNR-INM, National Research Council of Italy, 
Institute of Marine Engineering,\\Via di Vallerano 139, 00128 Roma, Italy}

\cortext[cor1]{Corresponding author}

\begin{abstract}
In this paper the hydrodynamics of fuselage models
representing the main body of three different types of aircraft,
moving in water at constant speed and fixed attitude is investigated 
using the Unsteady Reynolds-Averaged Navier-Stokes (URANS)
level-set flow solver $\chi$navis. The objective of the CFD study is
to give insight 
into the water landing phase of the aircraft emergency ditching. 
The pressure variations over the wetted
surface and the features of the free surface are analysed 
in detail, showing a marked difference among the three shapes 
in terms of the configuration of the 
thin spray generated
at the front.
Such a difference is a consequence of the different transverse
curvature of the fuselage bodies.
Furthermore, it is observed that at the rear, where a
change of longitudinal curvature occurs, a region of negative pressure 
(i.e. below the atmospheric value) develops. This 
generates a suction (downward) force of pure hydrodynamic origin.
In order to better understand the role played by the longitudinal 
curvature change on
the loads, a fourth fuselage shape truncated
at the rear is also considered in the study.
The forces acting on the fuselage
models are considered as composed of 
three terms: the viscous, the hydrodynamic and
the buoyancy contributions. For validation purposes the forces
derived from the numerical simulations are compared 
with experimental data.
\\
\\
\noindent $\copyright$ 2022. This manuscript version is made available under the CC-BY-NC-ND 4.0 license \url{http://creativecommons.org/licenses/by-nc-nd/4.0/}
\\
\noindent The final published journal article can be found at \url{https://doi.org/10.1016/j.ast.2022.107846}
\end{abstract}


\end{frontmatter}


\section{Introduction}
\label{sec:introduction}
%
%
Aircraft ditching, i.e the emergency water landing manoeuvre, although being a rare event, 
must be taken into account by the aircraft manufacturers in the certification process.
Ditching is a complicated manoeuvre that has to be
considered if, in presence of large damages or serious hazard, 
it is not possible to reach the closest runaway, see 
\cite{climent2006aircraft}. 
Given the uncertainty on the
weather and sea conditions and on the aircraft status, the trajectory of 
the aircraft at ditching and the effects of the water impact
cannot be easily predicted a priori. Therefore, the airworthiness regulations
provide recommendations for an aircraft design that must guarantee at most a
limited damage of the structure, in order to minimize the risk of injury of
the passengers and and allow enough floating time for 
a safe evacuation.

The whole ditching manoeuvre can be divided into different phases, among
which the most critical is surely the water impact phase, occurring
at a high horizontal speed. 
Owing to the costs of full scale tests, the water impact phase was typically 
investigated experimentally by considering small components 
tested in conditions resembling, as much as possible, the real ones.
Water impact tests of a flat plate at high horizontal speed were performed 
by \cite{smiley1951experimental}, who analysed the pressure distribution,
loads and wetted area for different impact conditions. \cite{naca2929} conducted 
free flight ditching tests of scaled fuselages with different shapes
to analyse the role played by the 
longitudinal and transversal curvature of the body on the
resulting  dynamics. 
{More recently, water entry tests on flat plate ditching at
high horizontal speed were performed in \cite{iafrati2015high} and 
in \cite{iafrati2016experimental}. Particular attention was also paid
to the analysis of the structural deformations, 
see \cite{spinosa2021experimental}. The study was further extended to 
double curvature specimens, for which hydrodynamic phenomena such as 
cavitation and ventilation have been observed to be important (see
\cite{iafrati2019cavitation,iafrati2020experimental}).
Over the years several numerical
methods have also been developed and validated against the experimental results,
as detailed in \cite{hughes2013aerospace,climent2006aircraft,bisagni2018modelling,
anghileri2011rigid,anghileri2014survey,
xiao2017development,duan2019numerical,woodgate2019simulation}.
These studies mainly focused on free-body water entry,
i.e with a body trajectory
that is not controlled and that freely evolves in time 
due to the hydrodynamic impact forces and gravity. 
These works have shown that during the impact phase a localised
high pressure area develops on the submerged part of the aircraft at
the spray root (\cite{climent2006aircraft,seddon2006review}). The
large pressure areas result in very large hydrodynamic
loads, which can lead to a possible structural damage of the
fuselage and to fluid-structure interaction phenomena 
(\cite{groenenboom2010hydrodynamics,cheng2011simulation,hughes2013aerospace,
groenenboom2015fluid,spinosa2021experimental}).

The successive ditching phase is typically referred to as the water landing phase,
during which the aircraft progressively adjusts its attitude and reduces its 
speed up to the final floatation phase. 
The dynamics of the aircraft during this phase and
the role played by the fuselage shape were investigated 
in \cite{wagner1948planing,naca2929,climent2006aircraft,hughes2013aerospace,Zhang2012}.
In particular, in the latter paper, the occurrence of suction 
forces at the rear and their effects on the aircraft dynamics were highlighted.
As shown in \cite{iafrati2019cavitation}, for a given fuselage shape
the forces at the rear vary substantially as a consequence of hydrodynamic
phenomena like cavitation and ventilation, which are highly sensitive to 
the speed, thus making the experiments at model scale not fully 
representative of the physics of the problem. 
Given the difficulty in setting up full scale experiments, 
computational approaches provide a good alternative, once 
they are carefully validated versus representative experiments.

For this purpose, within the H2020-SARAH project two different 
experimental campaigns were
performed, among others. The first one focused on the impact phase 
and on the effect of the fuselage shape and speed on the cavitation and ventilation
phenomena. The objective was to verify the capability of the computational 
models to correctly capture those hydrodynamic phenomena and their effects on 
the pressure and load distributions.
The second campaign concerned guided ditching tests at model scale, 
in which the whole trajectory of the body was prescribed.
These tests covered all the ditching phases, i.e the impact and landing phase.
The objective of these tests was to provide a dataset to assess
the capability and to validate the computational models 
to predict the loads and moments resulting from
the dynamics of the fuselage. In order to do so, guided tests
are preferred to free-body tests, since they guarantee a more precise 
control of the attitude and a much better repeatability. 
The rationale behind this choice is
that computational models can integrate quite accurately
the equations of the dynamics
of the body motion, provided that loads and moments
are correctly predicted.

In the latest phase of the guided ditching tests
the fuselage moves at constant speed and fixed attitude,
resembling a planing hull, and the aircraft reduces its speed
mainly under the action of the drag.
In order to understand the physics of the problem,
some specific tests at fixed attitude and speed were also performed.
These tests provide data over a time interval much longer than
the actual landing phase of a complete ditching test,
reducing the uncertainty of the measurements.  
Based on those experimental data, as a first step towards 
the development of high-fidelity computational models able to deal with
the ditching event, the planing motion of fuselage models in different conditions
is numerically simulated
by means of an Unsteady Reynolds Averaged Navier-Stokes
equations (URANS) flow solver. The numerical solver ($\chi$navis) is the same used in
\cite{broglia2018accurate}, \cite{iafrati2008hydrodynamics} and \cite{broglia2010hydrodynamics}
for the flow investigation around high-speed planing hulls. 
Simulations are performed on three different fuselage shapes, 
which are characterized by different curvatures in both the
longitudinal and transverse direction. Particular attention 
is paid to the role played by the fuselage shapes on the pressure 
distributions and loads. Comparisons with the experimental 
data are also established.
\section{Fuselage Models and Operative Conditions}
\label{sec:physical_shapes}
%
The studies are conducted on three different fuselage models, 
the shape of which is defined analytically, 
as detailed in \cite{iafrati2020experimental}.
The shapes are shown in \autoref{fig:Fuselage_Shapes}.
%
\begin{figure}[htb!]
  \centering
  \includegraphics[width=0.95\textwidth]{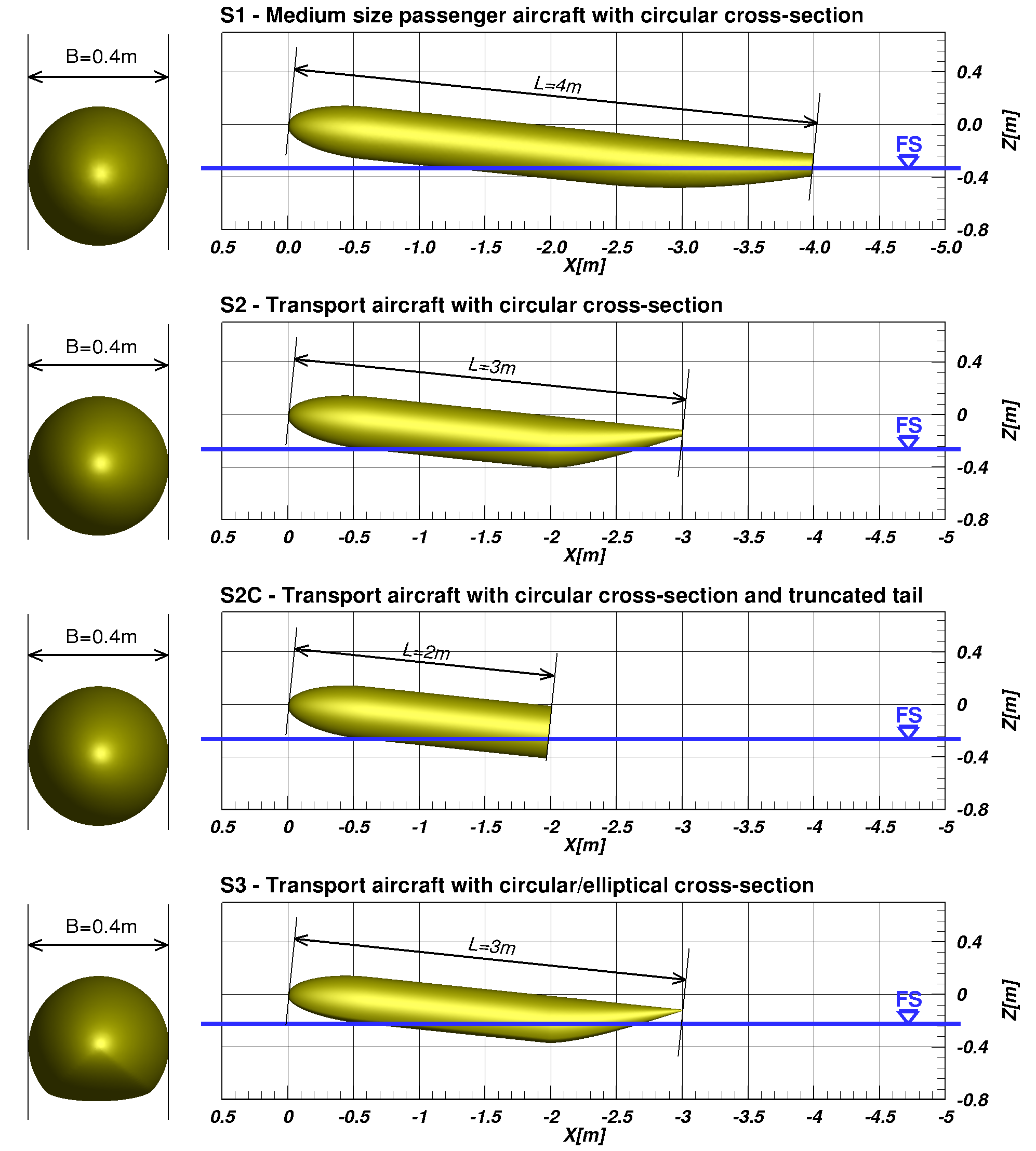}
  \caption{Fuselage shapes: side view (right column) and front view (left column).}
  \label{fig:Fuselage_Shapes}
\end{figure}
The fuselage shape \textbf{S1} resembles that of a small size passenger 
aircraft with a circular cross section,
whereas the shapes \textbf{S2} and \textbf{S3} have a longitudinal profile
similar to that of a freight transport aircraft,
\textbf{S2} being characterized by a circular cross section 
and \textbf{S3} by a circular-elliptical cross section.
Such a choice allows to investigate separately the effects of the longitudinal 
and transverse curvatures on the hydrodynamics and on the induced loads.
Behind the nose region the fuselages display a cylindrical shape with a constant
cross section. At the rear, the cross-section shrinks and moves up, 
leading to a bottom profile that varies rather smoothly in the shape
\textbf{S1} and more abruptly in the shapes \textbf{S2} and \textbf{S3}.
All the fuselage models have a breadth of 0.4~m, whereas the overall length varies 
between 3~m ({\bf S2} and {\bf S3}) and 4~m ({\bf S1}).
In the fuselages {\bf S2} and {\bf S3} the
cross section start shrinking from a distance
of 2~m behind the fuselage nose (going from left to right) 
whereas in the fuselage {\bf S1} from a distance of 2.4~m.
At those points a change in the longitudinal curvature occurs.
In the case of the  fuselage shape S1 
the shrinking of the cross-section is much smoother,
resulting in a milder longitudinal curvature, 
see again \autoref{fig:Fuselage_Shapes}.
As a further study, a more extreme change in the longitudinal
curvature is investigated
in shape {\bf S2C}, which is truncated at a distance of
2~m behind the nose.
As detailed below, experimental measurements of hydrodynamic 
loads are available for \textbf{S1}, \textbf{S2} and \textbf{S3},
whereas {\bf S2C} has been only simulated.

In the following, a body fixed frame of reference,
with the origin located on the fuselage nose 
is considered (\autoref{fig:Ref_Frames}). The coordinates are indicated with capital letters 
$X$, $Y$ and $Z$. The $Z$ axis is aligned with the
gravity acceleration $\bf {g}$, but positive upwards, the $X$ axis is
aligned with the velocity of the towing carriage, pointing ahead. The $Y$
axis completes a right-handed coordinate system. The $X$-$Z$ plane, also
denoted as the mid-plane in the following, is parallel
to the undisturbed free surface.
%
\begin{figure}[htbp]
	\centering
	\includegraphics[width=1.00\textwidth]{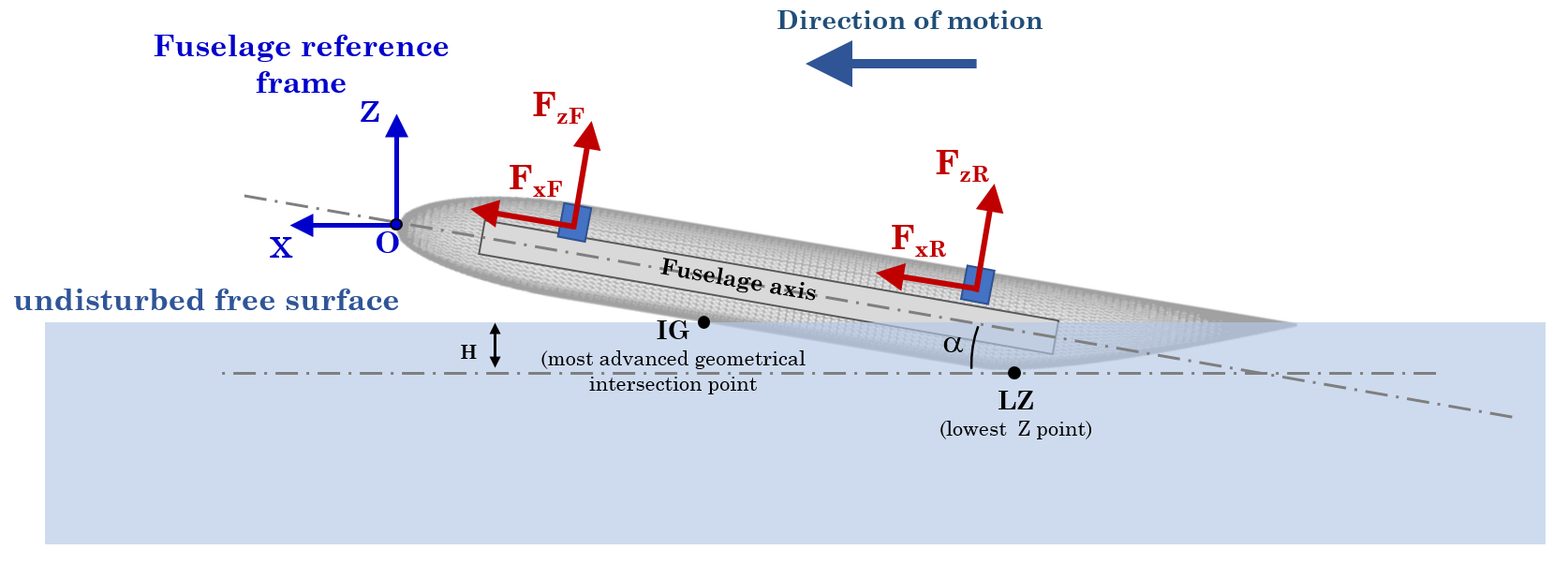}
	\caption{Reference frame used for the numerical data analysis and
  directions/signs of the forces recorded by the load cells.}
	\label{fig:Ref_Frames}
\end{figure}
The immersion depth of the fuselages $H$ is defined as the distance 
from the undisturbed free surface and {\bf LZ}, i.e the point of
the model at the lowest $Z$-coordinate at the specific
pitch angle of the simulation, as shown in \autoref{fig:Ref_Frames}.
Instead, the most advanced geometrical intersection point between 
the lower longitudinal profile of the fuselage and the undisturbed 
water level is indicated by \textbf{IG}.
Both these points are located in the mid-plane. 

The numerical simulations are performed
in deep fresh water conditions at a temperature of 16$^{\circ}$, at which
the density is $\rho=998.9$~kg/m$^3$ and the kinematic viscosity is
$\nu =1.107 \times 10^{-6}$~m$^2$/s. The fuselage advances 
at a speed of $U=5$~m/s 
with a fixed pitch angle of $6\degree$ and a nominal immersion depth 
of $H_{num}=146$~mm. 
The reference length and velocity to define
the non-dimensional quantities
and numbers are the length of the fuselages {\bf S2} and {\bf S3}, i.e. $L=3$~m
and the towing speed $U$, respectively.}
The corresponding Reynolds and Froude numbers are
$Re=U L/\nu=1.355 \cdot 10^7$ and $Fr=U/\sqrt{g L}=0.922$,
since the acceleration of gravity is assumed to be $g=9.81$m/s$^2$.
It is worth noting that during the corresponding experiments the actual 
immersion depth is different due to the lowering of the free surface associated to
the aerodynamic effect of the carriage, as explained in Section \ref{sec:expsetup}.
%
\section{Numerical Setup}
\label{sec:MathMod}

\subsection{Mathematical and Numerical Models}
The numerical simulations are conducted by means of the in-house developed
$\chi$navis code, solving the URANS equations. The solver is based on 
a finite volume discretization of
the conservative equations for an incompressible viscous flows, which read as:
\begin{equation}
\label{eq:NSint}
\ds{{\bf \Lambda}\dert{ } \int_{\calV} {\q} \diff V +
  \int_{\calA(\calV)}
  [\Phi^c(\q) - \Phi^d(\q)] \diff S = {\bf 0}},
\end{equation}
having denoted with $\calV$ a control volume and with
$\calA(\calV)$ its the boundary. $\diff S$ and $\diff V$
are the infinitesimal volume and surface element, respectively.
In this formulation all the variables are expressed in their
non-dimensional form.
In \autoref{eq:NSint}, $t$ is the time, $x_i$ is the i-th space coordinate,
${\bf \Lambda}$$=$$\mbox{diag} \left( 0, 1, 1, 1 \right)$
is a diagonal matrix, and
$\q=(p,u_1,u_2,u_3)^T$ is the vector of the non-dimensional
variables for incompressible
flows, which are the velocity components $u_i$
along the three directions and the hydrodynamic pressure $p$,
which is related to the fluid pressure $\hat{p}$ and the acceleration 
of gravity $\bf g$
by $p$$=\hat{p}$$+(z-z_{FS})/Fr^2$, being $z_{FS}$ the (undisturbed)
free surface level. It is worth noting that in the following the hydrodynamic pressure $p$
is the one used in the data analysis and in the graphs.
$\Phi^c$ and $\Phi^d$ are the convective and viscous fluxes, i.e.
$\Phi^c = [ u_i n_i ; u_j u_i n_i + p n_j ]^T$ and $\Phi^d = \left( 0 ; \tau_{ji} n_i \right)^T$,
where $i,j=1,2,3$. $n_i$ is the $i$-th component of the outward unit vector normal
to the surface $\calA(\calV)$
and 
$\tau_{ij} = \left(\nu_t + 1/Re \right) \left({\partial u_i}/{\partial x_j} + {\partial u_j}/{\partial x_i}\right)$
are the components of the stress tensor.
Finally, $\nu_t$ is the turbulent viscosity, 
which is modelled via the \cite{spall} model.

The problem is closed by enforcing appropriate conditions at physical and
computational boundaries. On solid walls, the relative velocity is set to
zero (whereas no condition on the pressure is required). At the (fictitious)
inflow boundary, the velocity is set to the undisturbed flow value, and the
pressure is extrapolated from inside. On the contrary, the pressure is set
to zero at the outflow, whereas velocity is extrapolated from the inner points.
At the inlet, outlet and symmetry boundaries 
the extrapolation of pressure and velocity 
is obtained using a second order accurate approximation of their normal derivatives. 
For the conformal adjacent blocks, no interpolation is used; the values 
in the adjacent blocks are simply stored in a double layer of ghost cells. 
As for the overlapping grids, a tri-linear interpolation is used.

At the free surface, whose location is one of the unknowns of the problem, 
the continuity of stresses across the surface is required. If the presence of 
the air is neglected, the dynamic boundary condition reads:
\begin{equation}
\begin{array}{l}
\ds{p = {\tau _{ij}}{n_i}{n_j} + \frac{z-z_{FS}}{{\mbox{Fr}^2}}
}\\[2mm]
{\tau _{ij}}{n_i}t_j^1 = 0\\[2mm]
{\tau _{ij}}{n_i}t_j^2 = 0
\end{array}
\label{eq:freesurf-bc}
\end{equation} 
where
{\mbox{\boldmath $n$}},
{\mbox{\boldmath $t^1$}} and
{\mbox{\boldmath $t^2$}}
are the surface normal and the two tangential unit vectors, respectively. 
Surface tension effects are not taken into account.

The free surface, which has equation $\eta(x,y,z,t)=0$, is computed from the kinematic condition:
\begin{equation}
\frac{D\eta(x,y,z,t)}{Dt}=0
\label{eq:freesurf-pos}
\end{equation} 
Initial conditions have to be specified for the velocity field and for the free surface:
\begin{equation}
\begin{array}{l}
{u_i}\left( {x,y,z,0} \right) = u_i^0(x,y,z)\quad i = 1,2,3\\
\eta(x,y,z,0) = {\eta^0}(x,y,z)
\end{array}
\label{eq:init-cond}
\end{equation}

The numerical solution of \autoref{eq:NSint} is obtained
on a computational domain discretized by adjacent or overlapped structured blocks
composed of hexahedral volumes. A suitable overlapping grids approach is exploited
to handle either complex geometries or multiple bodies in
relative motion (see \cite{Zaghi2014}).
Conservation laws are enforced in each control volume
in an integral form. The surface integrals are evaluated by means of a second-order formula.
For the viscous fluxes, the velocity gradients
are computed using a standard second-order centred finite volume
approximation. Numerical convective fluxes are computed solving a
Riemann problem whose left and right states are estimated by means of
a third-order upwind-biased scheme \citep{vanleer}.
A second-order accurate solution of the Riemann problem
is derived, which reduces
the computational cost necessary to compute iteratively the exact
one (see \citealt{PosaBroglia_IB2019} for more details).
Time integration is performed by a second-order three-points
implicit backward finite-difference scheme.
The resulting system of coupled non-linear algebraic equations is tackled via a
{\it pseudo-time} technique (\cite{chorin}), using an implicit Euler scheme
with an approximate factorization (\citealt{beamwarming}).
A local dual time step and a multi-grid technique
are exploited for faster convergence in the pseudo-time.

Free surface effects are taken into account by a single phase level-set
algorithm \citep{LS_CaF}, \citep{broglia2018accurate}.
The resulting scheme is globally second-order accurate
in both space and time,
producing oscillation-free solutions, also in presence of discontinuities
(see, for example,
\cite{DiMascio200919}).
\subsection{Fluid Domain and Computational Grid}
\label{sec:comp_grid}
Taking advantage of the symmetry of the flow with respect to the mid-plane
$X-Z$, only half of the domain is considered. In the $X$-direction the domain
extends $1.5L$ upstream and
$4L$ downstream of the fuselage nose, i.e. the computational domain is
$5.5L$ long. In the $Y$-direction it extends from the symmetry
plane to $Y=2L$, whereas in the $Z$-direction from $0.2L$ above the fuselage
nose to $2L$ below it.

\begin{figure}[htb]
\centering
\includegraphics[width=0.75\textwidth]{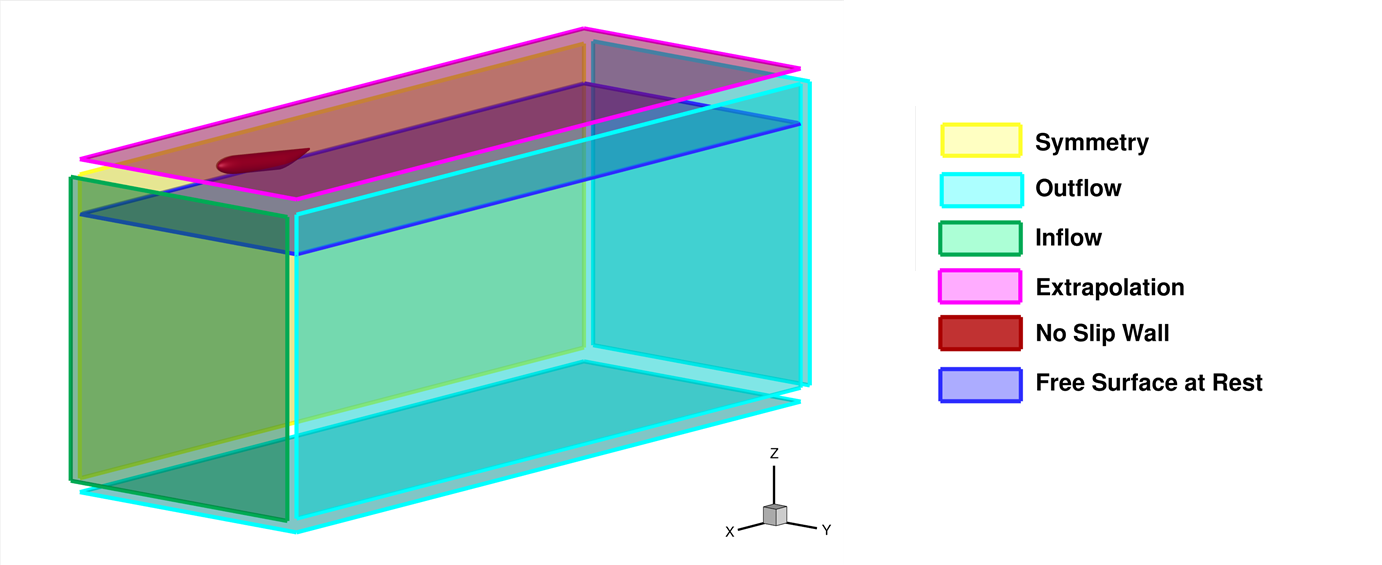}
\caption{Computational domain and boundary conditions.}
\label{fig:compdom}
\end{figure}

The boundary conditions are summarized in (\autoref{fig:compdom}) .
Symmetry boundary conditions are imposed on the $Y=0$ plane. 
It is worth noting that, even if the 
top boundary is placed only at $Z=0.2L$ above the fuselage nose,
the air flow is not simulated, thus this circumstance does not 
affect the validity of the simulations. At this boundary all the variables
are simply extrapolated.

%
%
\begin{figure}[htb]
\centering
\includegraphics[width=0.715\textwidth]{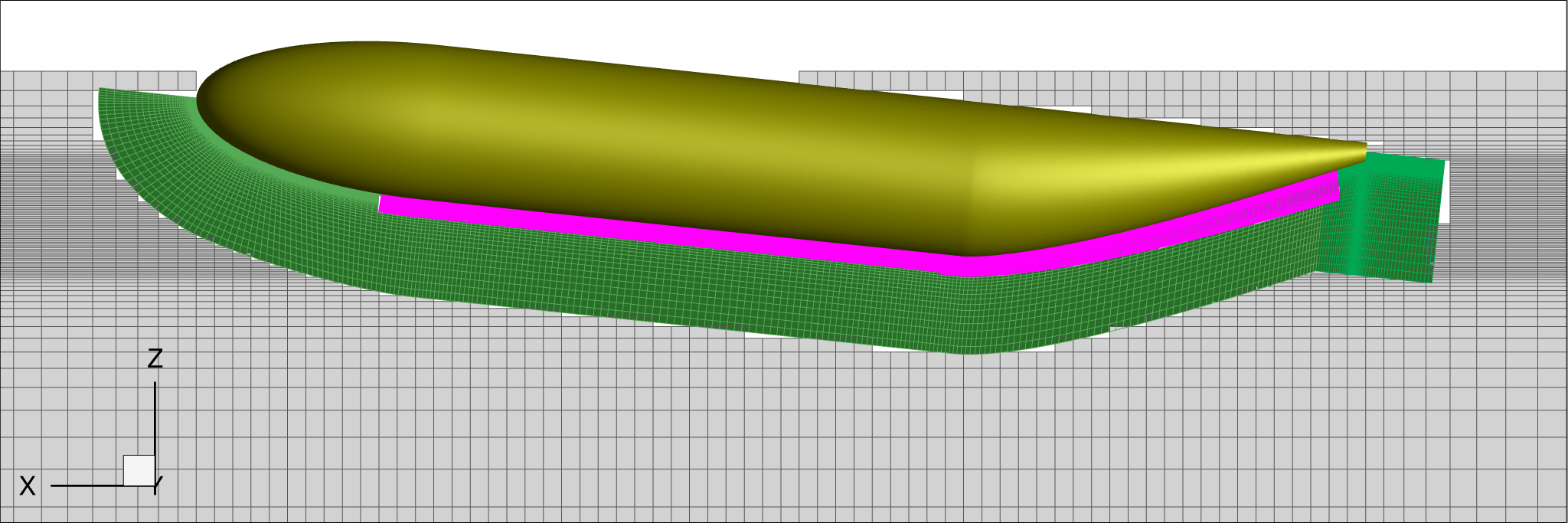}
\mbox{\ \ }
\includegraphics[width=0.245\textwidth]{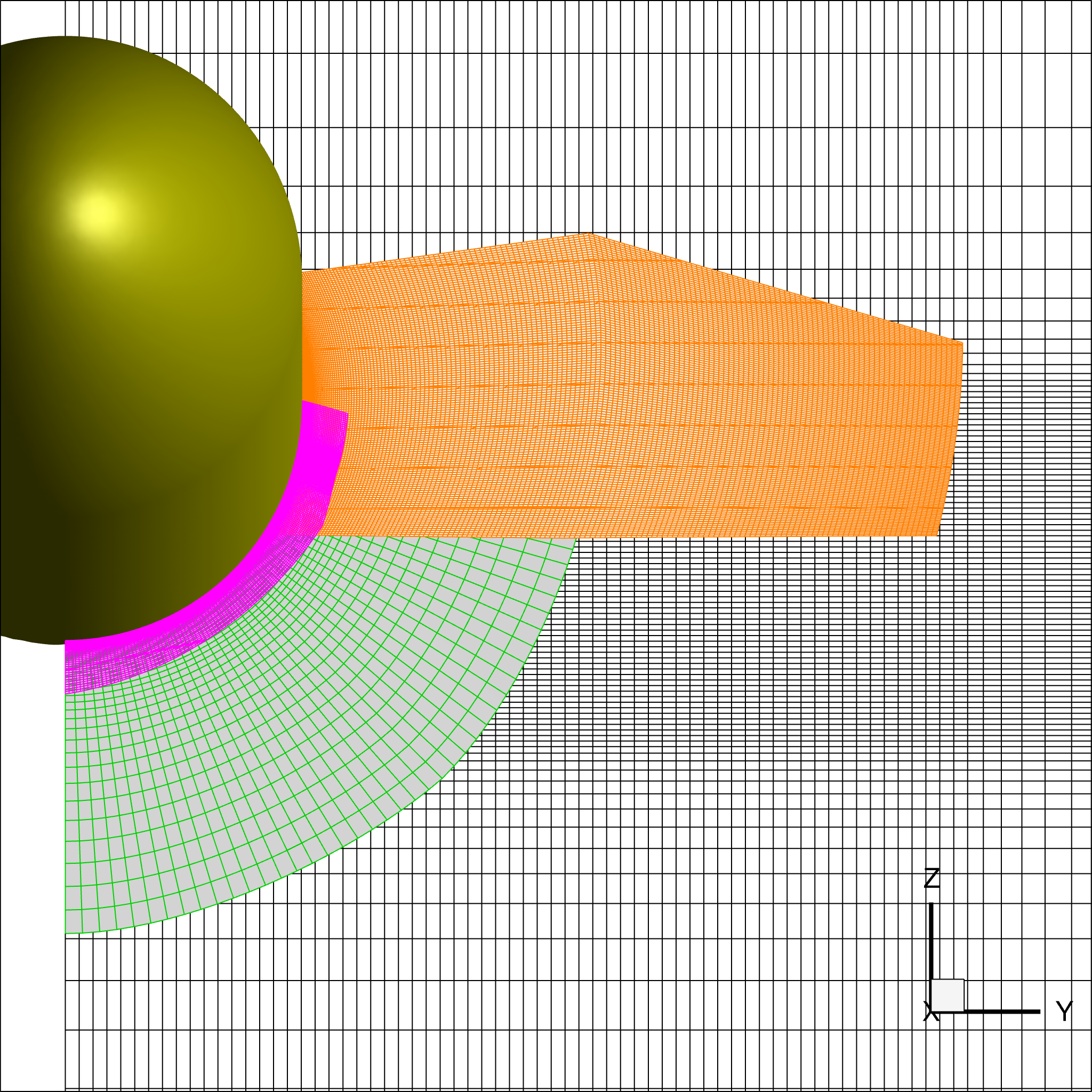}
\caption{Details of the computational mesh.}
\label{fig:mesh}
\end{figure}

The computational mesh is a structured grid composed
of adjacent and overlapping blocks. As shown in the detailed views 
shown in \autoref{fig:mesh}, 
the near wake region (in green) is further refined (in purple)
to capture accurately the formation of the spray root.
A refined region is also considered farther away from the wall (in orange) 
to capture the 
plunging of the spray detaching from the fuselage on the free surface.

The height of the first cells of the mesh on the surface is set in order to
achieve an average wall normal spacing below 1 viscous unit.
The outer part of the domain is discretized by a Cartesian background mesh.
A similar grid topology is
adopted for the different test cases, whereas the block 
distribution and grid 
spacing are tailored for the specific case. The total 
number of discretization
volumes depends on the test case, and counts up to about 15 millions 
of control volumes.

The numerical solutions are computed on four grid levels,
generated by removing every other grid points from the next finer one.
In the grid sequencing procedure,
the solution is first computed on the coarsest grid level, then, it is
prolonged onto the next finer grid. On the finer meshes, the multi-grid
algorithm accelerates the convergence in the pseudo-time
using all the available coarser grid levels. A physical non-dimensional time
step equal to $5\times10^{-3}$ is used for the time integration on the
finest mesh, while on coarser mesh the time step is multiplied by a factor 2.

The computational problem is tackled here via parallel High Performance
Computing, distributing the structured blocks among all available CPUs.
Communications across them for the coarse grain (distributed memory)
parallelization is achieved via calls to standard Message Passing Interface
(MPI) libraries, whereas for fine grain (shared memory) parallelization is based
on Open Message Passing (OpenMP) libraries.
\section{Experimental setup}
\label{sec:expsetup}
%
The experiments are carried out in the CNR-INM Wave Tank.
The towing carriage is operated at a speed of
5 m/s, with a maximum fluctuation of about 0.15~\%.

The tests are carried out in calm water and at fixed attitude (\emph{captive tests}), 
i.e. at a fixed pitch angle $\alpha = 6^{\circ}$
and immersion depth $H_{exp}$=150~mm.

\begin{figure}[htbp]
	\centering
		\includegraphics[width=0.90\textwidth]{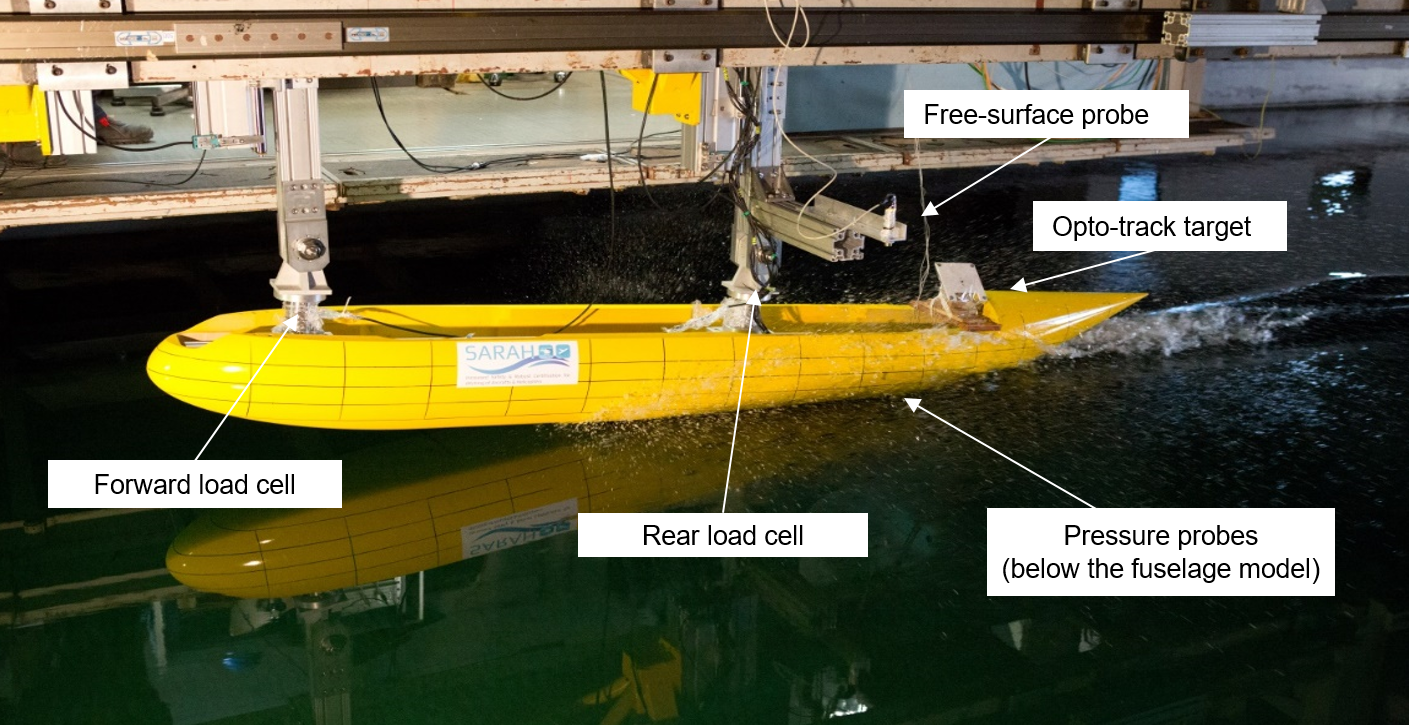}
	\caption{Picture of the instrumented fuselage model {\bf S3}.}
	\label{fig:exp_setup_S2}
\end{figure}

The attitude regulation is performed through a 2-degree-of-freedom system
driven by two linear servo-actuators. The instrumented fuselage
model {\bf S3}, installed on the carriage is shown in \autoref{fig:exp_setup_S2}.
%
The free surface probe located at the rear of the fuselage is used for the
positioning of the fuselage with respect to the undisturbed water level before
each run. The opto-track system is employed as a cross-check
of the free surface probe measurements.
Thanks to the combined use of these devices,
the final position can be considered quite accurate, with an uncertainty of
$\pm$1~mm. The connection of the fuselage models to the actuator system
is achieved by means of two 6-axes piezoelectric load cells.
The load directions and signs are shown in \autoref{fig:Ref_Frames}. The forces
$F_{xF}$ and $F_{xR}$, referring to the front and rear load cells respectively, 
are parallel
to the fuselage axis, whereas $F_{zF}$ and $F_{zR}$
are normal to the fuselage axis in the vertical plane.

For their operational principle, the load cell readings are zeroed before
any acquisition. The zeroing is performed over a short time interval (about 1~s)
when the fuselage models are already submerged in water 
at the required attitude. In this position
the forces acting on the fuselage are the weight and the hydrostatic
buoyancy.

Although this type of \emph{captive tests} guarantees a very 
precise control of the attitude and of the immersion depth of the model,
they are characterised by some drawbacks that have to be accounted for.
First of 
all the, the towing carriage in use is located very close to the water surface
of the basin. This creates
an unavoidable aerodynamic effect, for which the water level below
the carriage during the run is slightly different than that at rest. 
Through some specific measurements of the free surface level performed 
without the model installed, it is found that the free surface lowering
at a carriage speed of 5~m/s is about 4~mm. This is the 
reason why, to facilitate the comparison of the results,
the numerical simulations are performed at $H_{num}$=146~mm.
Secondly, after each run some residual waves remain,
which are characterized by a very low decay rate. 
Even after thirty minutes, a standing wave
with a period of 70~s and amplitude of about 3~mm is observed.
The only way to remove the effects of the residual standing wave
on the force measurements is to perform averages
over sufficiently long time intervals.

Finally, it is worth noting that
while the present present simulations
are performed at fixed attitude, in the aircraft ditching case the 
vertical and horizontal velocities are the results of the combined effects of the
hydrodynamic loads, the aerodynamic loads, determined by the aircraft wings and control
surfaces, and by the action of the engines. As such, the actual loads 
may differ from what found in the present simulations.
%
%
\section{Results}
\label{sec:results}
The flow field characteristics are first described in details
for the fuselage shape {\bf S2}. In order
to explain more deeply the physical origin of the suction area at the
rear, where the longitudinal curvature change occurs,  
the surface pressure field of {\bf S2} is compared with that of {\bf S2C}, 
i.e. the truncated fuselage model. 
Hence, the effects of the fuselage shape in terms of pressure and 
free surface shape at the spray root and at the rear are analysed. 
Successively, the overall forces $F_X$ and $F_Z$ acting on the fuselage models, 
as well as their longitudinal distributions
are derived. The
forces are investigated by decomposing them into three contributions:
the viscous, the buoyancy and the hydrodynamic ones.
The forces computed by the numerical simulations are finally compared 
with the experimental measurements for the purpose of validation.
\subsection{Spray evolution and wave pattern}
\label{sec:jetshape}
\begin{figure}[htb]
	\centering
		\includegraphics[width=0.48\textwidth]{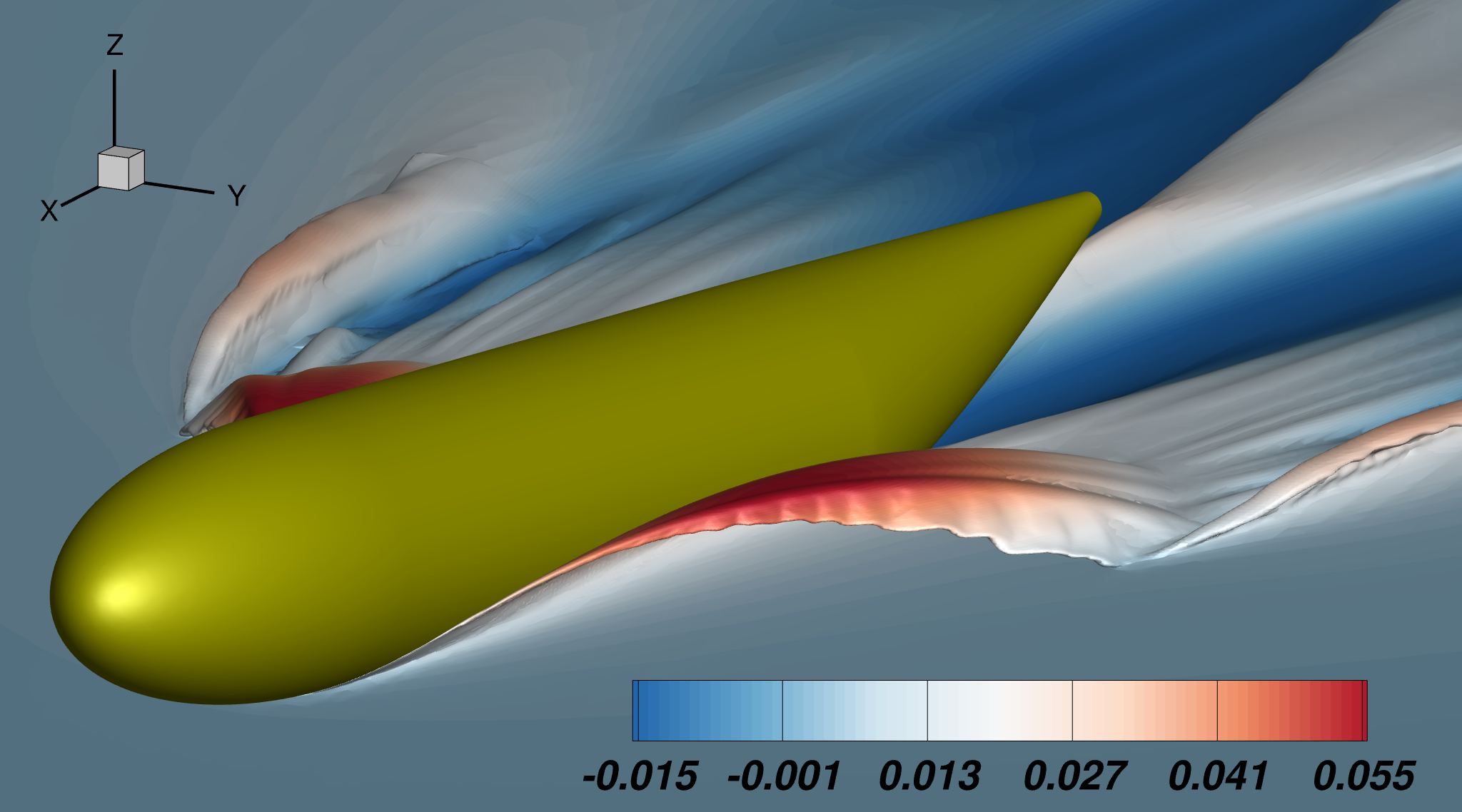}
		\includegraphics[width=0.48\textwidth]{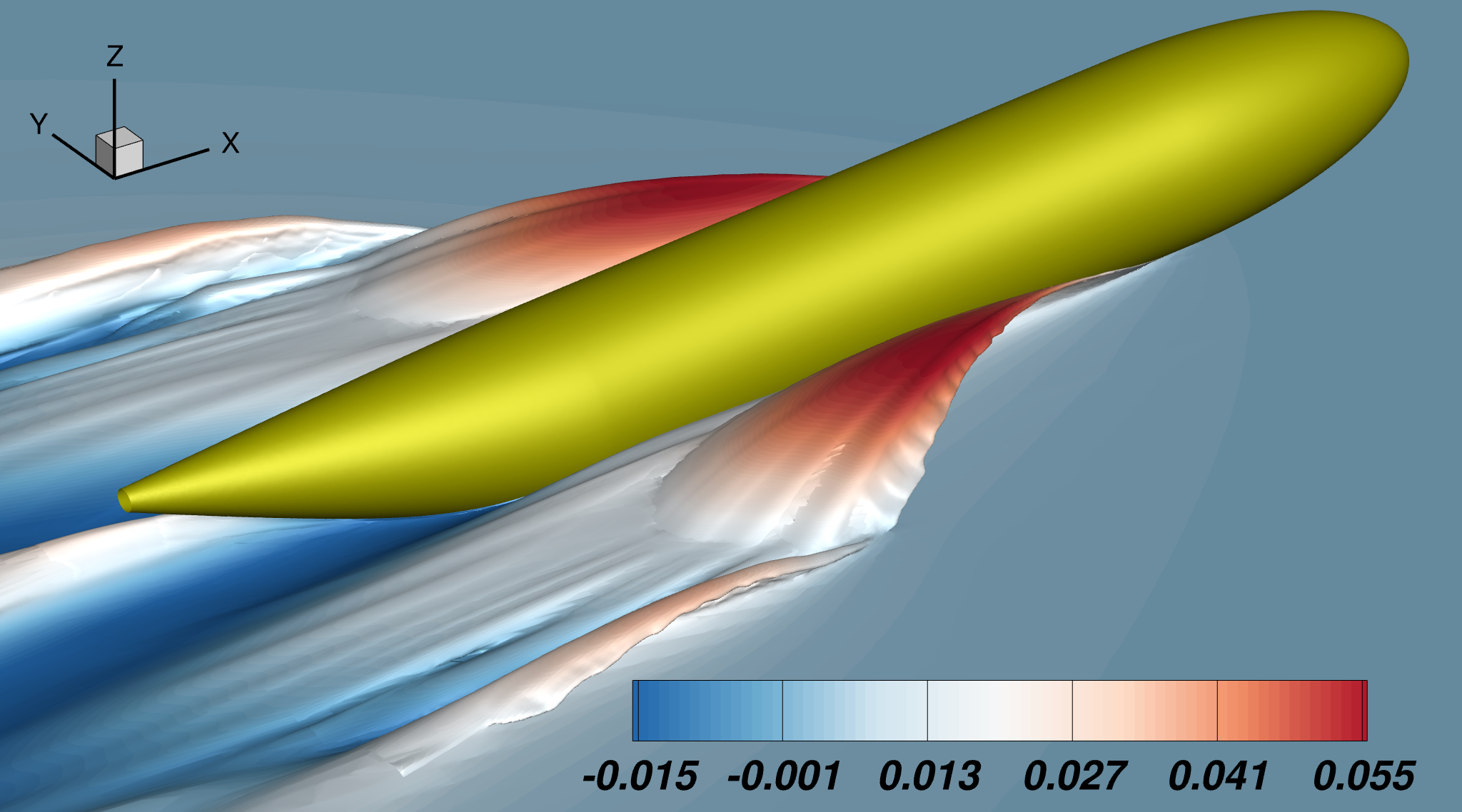}\\[2mm]
		\includegraphics[width=0.96\textwidth]{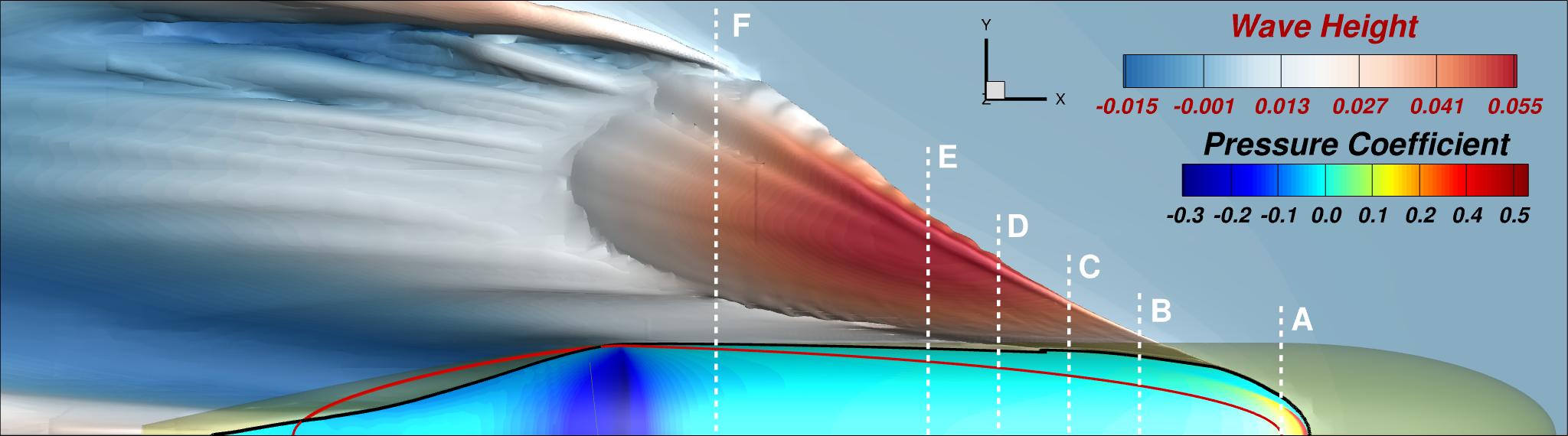}
	\caption{Free surface derived from the numerical simulation of 
the fuselage model {\bf S2}, view from the front and from the back (top)
 and view from above (bottom). 
In the bottom figure the pressure coefficient $C_p$  on the wetted area is also shown.
 Dashed lines and capital bold letters refer to the cross sections shown in the following.}
	\label{fig:freesurface_M2_5ms_6deg_146mm}
\end{figure}

An overview of the free surface shape around the fuselage and
of the pressure coefficient, defined as $C_p \doteq 2 p$, on the wall is shown in
\autoref{fig:freesurface_M2_5ms_6deg_146mm}, where, for the sake of clearness,
the solution is mirrored a the longitudinal plane of symmetry. The wave
pattern shown in this figure is that of the {\bf S2} shape, advancing at $U=5$~m/s,
with 6$^\circ$ pitch angle and an immersion depth of 146~mm. 
%
%
The main features of the flow field resemble those of planning hulls
(see for example \cite{broglia2018accurate}). Differently
from the displacement vessels case at low Froude numbers, 
in which a classical Kelvin wave system forms immediately
around the hull, in this case a spray is formed beneath the fuselage.
The spray leaves the fuselage and plunges onto the free surface,
leading to a rather complex flow behind. 
Due to the high pressure occurring
in the front part of the fuselage (see the panel at the bottom of
\autoref{fig:freesurface_M2_5ms_6deg_146mm}), this spray rises up
both ahead and laterally. 

The free-surface shape generated by the fuselage can also
be investigated by means of a 2D+$t$ approach
e.g. \cite{iafrati2008hydrodynamics,broglia2010hydrodynamics}. 
By doing so, the steady flow can be approximated as a water entry
problem in the transverse plane, on an earth fixed reference frame.
From this point of view, the spray remains attached to the body in an early stage
but detaches from the body afterwards, e.g. \cite{sun2006water}.

\begin{figure}[htbp]
	\centering
		\includegraphics[width=0.95\textwidth]{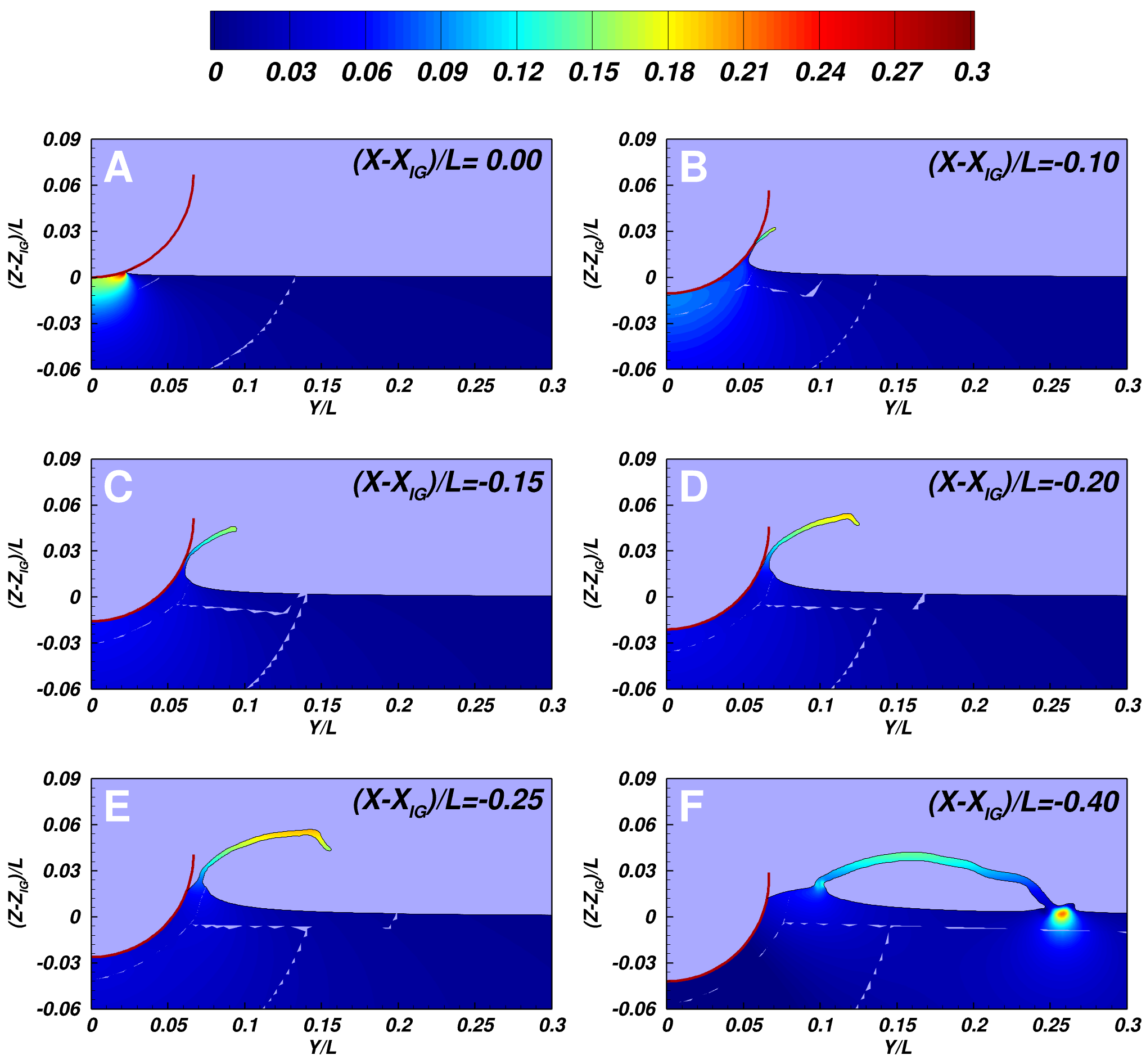}
	\caption{Free surface shape evolution along the $X$-axis of the fuselage \textbf{S2},
	coloured with the contour maps of the hydrodynamic pressure coefficient $C_p$. The red
	solid line represents the edge of the fuselage cross section at each
	specific $X$-location.}
	\label{fig:test_S2_slices_DX}
\end{figure}

This process is shown in more detail in \autoref{fig:test_S2_slices_DX}, 
where the pressure coefficient variation and the evolution of the free surface
are shown at different cross-sections in the longitudinal direction.
%
%
For convenience, the position of these cross-sections is relative to
$X_{IG}$, the abscissa of {\bf IG}.
Their positions are indicated with white dashed lines
in \autoref{fig:freesurface_M2_5ms_6deg_146mm}.
In the first cross-section (top-left)
the pressure peak at the spray root and the water rise can be noted.
The jet separation and its lateral evolution appear in the following two
cross-sections.
Moving backwards, the spray detaches at about  $(X-X_{IG})/L=-0.10$ and starts to plunge 
at $(X-X_{IG})/L=-0.25$. At $(X-X_{IG})/L=-0.40$ 
it falls onto the free surface, thus leading to a high pressure pulse.
As a consequence, see \autoref{fig:freesurface_M2_5ms_6deg_146mm}, 
a large spray sheet, hydraulic bores and ricochets form in the wave pattern. 
A strong interaction among these structures is observed, 
resulting in a very complex free surface shape in that region.

The lowering of the free-surface observed at the rear of
the fuselage (evident in the bottom
panel of \autoref{fig:freesurface_M2_5ms_6deg_146mm}), is a consequence
of the low hydrodynamic pressure region generated by 
the change in the longitudinal curvature that induces an increase 
in the fluid velocity.

Finally, behind the fuselage the accelerated flow leaves 
the body surface tangentially, leading to the formation of the typical 
rooster tail shape,
similar to that observed in high speed crafts.
\subsection{Hydrodynamic pressure on the fuselage}
\label{sec:hydro_pressure}
The contour maps of the hydrodynamic pressure coefficient
over the wetted surface are shown in the bottom panel
of \autoref{fig:freesurface_M2_5ms_6deg_146mm} and in 
\autoref{fig:comparison_pressure_M2_M2cut}. 

In the spray root region, just like in planing hulls
or in flat plates moving in water at high speeds (\cite{faltinsen2005hydrodynamics},
\cite{savitsky1964hydrodynamic}, \cite{iafrati2016experimental}, \cite{broglia2018accurate})
a region of maximum pressure is observed.
In this context \emph{flow separation} refers to
the detachment of the free surface from a solid wall. Accordingly, 
the \emph{separation line} is the contact line between them.
The separation line from the fuselage wall in motion is
shown in the bottom panel of \autoref{fig:freesurface_M2_5ms_6deg_146mm} 
with a black line, whereas the red thick line indicates the geometrical intersection 
line between the fuselage and the undisturbed free surface.
The separation line is located outside the geometrical intersection line,
this implying a a water pile-up,
and the distance between these lines increases while moving backwards
along the $X$-direction up to a certain point. 
Going backwards, the separation line gets closer the geometrical 
intersection line, until they meet at about $X=X_{LZ}$.
Behind \textbf{LZ}, due to the acceleration of the fluid and 
the lowering of the free-surface, the separation line is located 
inside the geometrical intersection line,
see again \autoref{fig:freesurface_M2_5ms_6deg_146mm}. 

It is worth noticing that the flow accelerates just ahead of
the curvature change ad decelerates 
behind it, i.e. moving from left
to right in~\autoref{fig:comparison_pressure_M2_M2cut}. 
This circumstance is associated with a pressure
drop to negative pressure values ahead the curvature change, 
followed by a pressure recovery behind.  
This situation can also be interpreted 
from a 2D+$t$ perspective (\cite{iafrati2008hydrodynamics}, \cite{broglia2010hydrodynamics}). 
From this point of view, the longitudinal curvature of the fuselage 
introduces a variation in the vertical entry velocity $V_b$ of the body contour 
in an earth-fixed transverse plane, which can be written as $V_b = V \sin \theta$, 
where $\theta$ 
is the local inclination angle of the bottom profile in the longitudinal plane. 
In the front part $\theta$ is constant and positive, and so is the vertical 
entry velocity. 
However, starting from the curvature change and going backwards, 
$\theta$ decreases and becomes progressively
negative, thus turning the problem from a water-entry to a water-exit one.
As shown in \cite{del2021water}, this causes the shrinking of the wetted
surface and the generation of negative pressure and suction loads.
\subsection{Comparison between the full and the truncated fuselage flows}
\begin{figure}[htb!]
	\centering
	\includegraphics[width=0.95\textwidth]{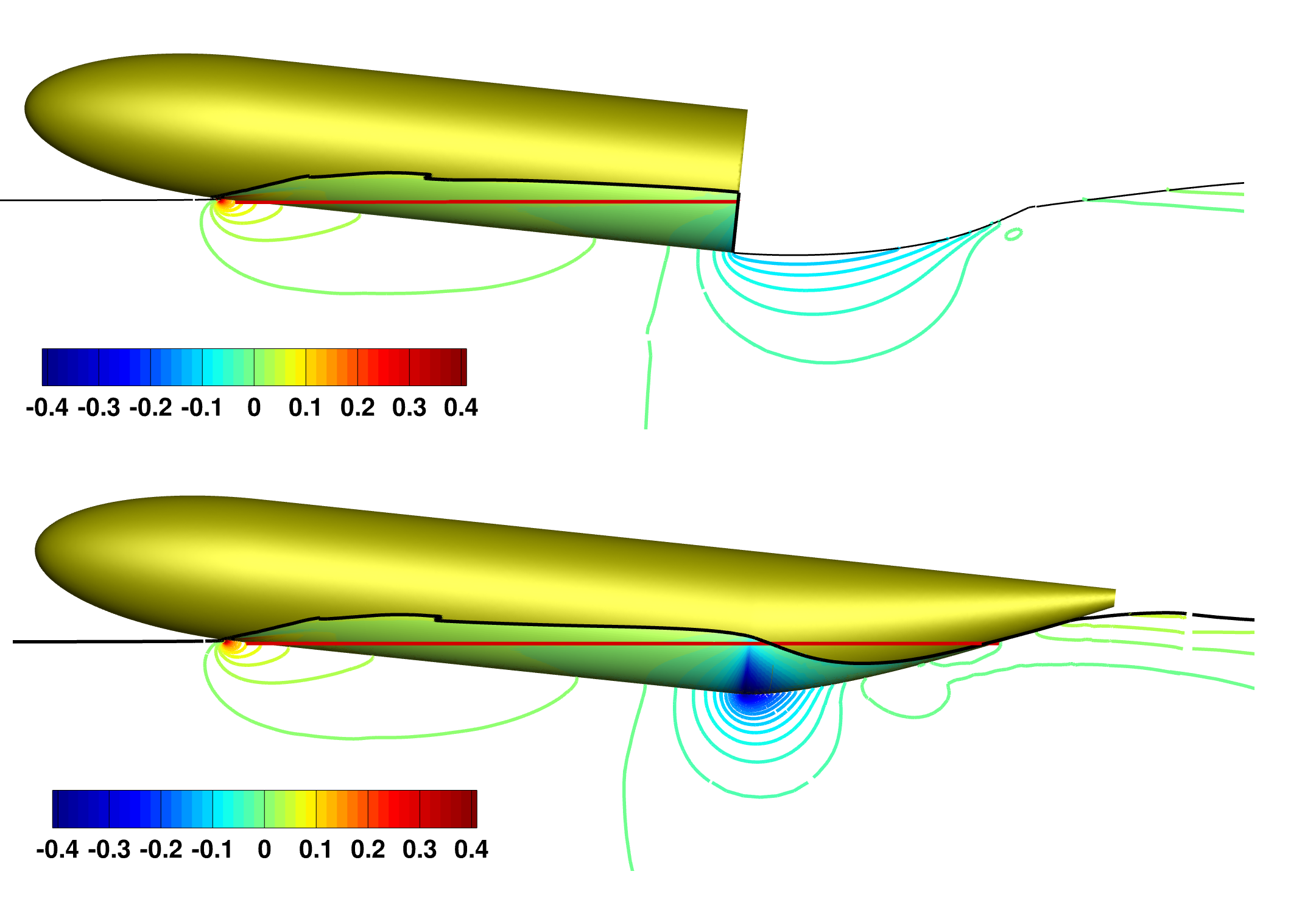}
	\caption{$C_p$ contour lines for the truncated fuselage \textbf{S2C} (top) and for the
	full fuselage \textbf{S2} (bottom) on the midplane $X-Z$ and $C_p$ colour contour maps over
	the wetted surfaces.}
	\label{fig:comparison_pressure_M2_M2cut}
\end{figure}

As mentioned above, in order to improve the understanding of the 
mechanism of pressure decrease at 
the curvature change at the rear, it is very informative to compare the simulation 
of the full fuselage shape {\bf S2} with that of the truncated fuselage
{\bf S2C}, shown in \autoref{fig:Fuselage_Shapes}, at the
same conditions.
A comparison between the $C_p$ contour lines in the mid-plane 
for \textbf{S2C} and \textbf{S2}, as well as between
the $C_p$ contour maps on the wall, 
is shown in \autoref{fig:comparison_pressure_M2_M2cut}.
%
As expected, the pressure fields in the front part present a very similar trend.
Instead, the $C_p$ pattern at the stern of {\bf S2C} resembles that of 
the dry-transom flow, typical of planing hulls, in which the free surface separation
from the body occurs exactly at \textbf{LZ}, i.e. at the transom, (see for example
\cite{broglia2018accurate,iafrati2008hydrodynamics,broglia2010hydrodynamics}),
which remains completely dry, with a hydraulic jump behind it.
The flow accelerates in the rear part of the truncated fuselage,
causing a weak pressure drop.
However, this pressure drop is much lower with respect to the full
fuselage case, since for the truncated case at {\bf LZ} the global pressure 
is forced to be almost null, due to the boundary condition at the free surface,
see \autoref{eq:freesurf-bc}.
On the other hand, as it is shown in \autoref{fig:comparison_pressure_M2_M2cut},
for the full fuselage shape {\bf S2}, the free surface
remains attached to the wall also behind {\bf LZ},
and a more intense flow acceleration results, leading to
the situation described in Section \ref{sec:jetshape} 
and Section \ref{sec:hydro_pressure}.
\subsection{Effect of the fuselage shape}
\begin{figure}[htpbp]
   \centering
   \includegraphics[width=0.99\textwidth]{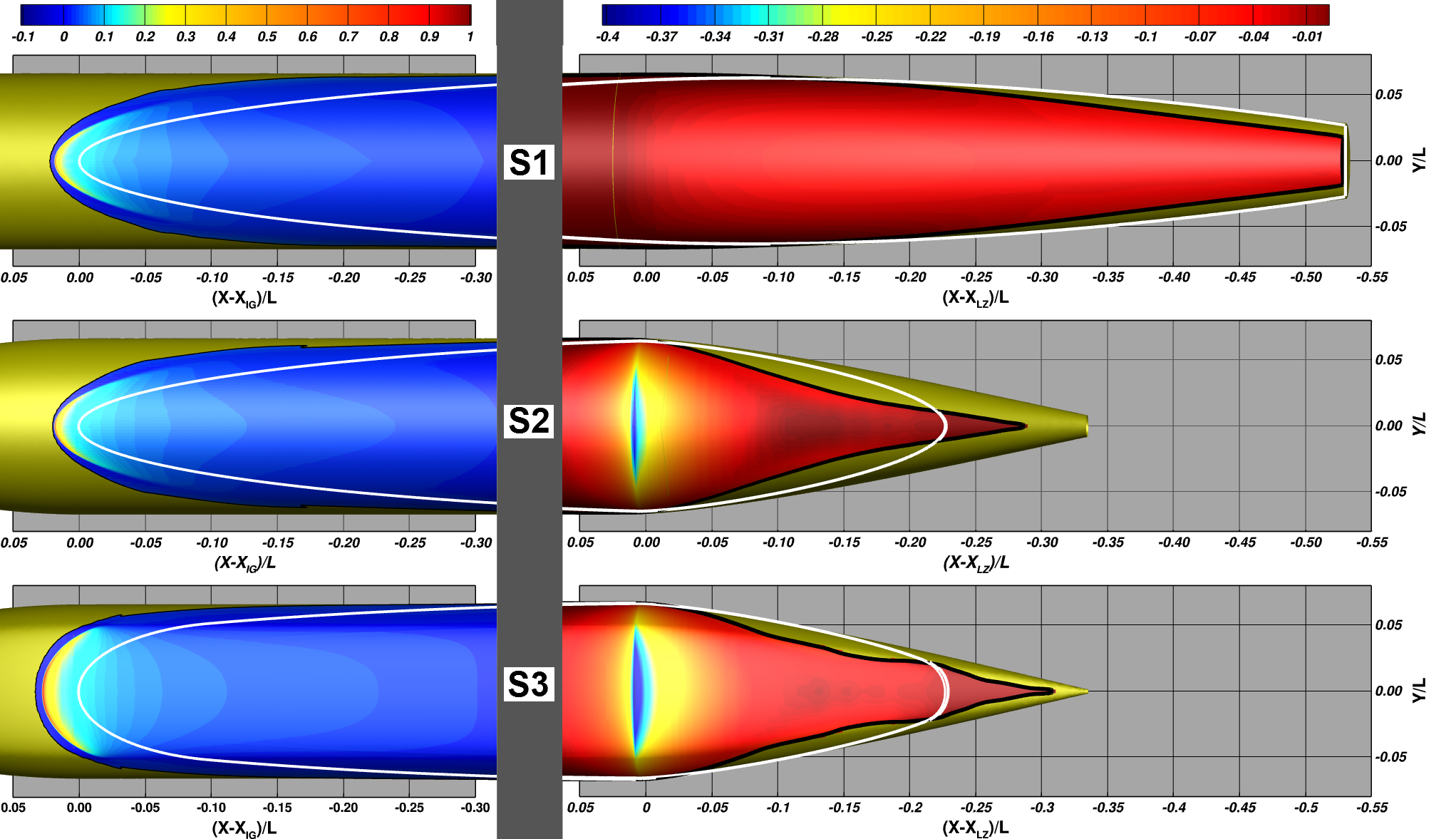}
   \caption{$C_p$ contour maps for the three fuselage shapes {\bf S1} (top),
	{\bf S2} (middle) and {\bf S3} (bottom), at the spray root (left) 
	and in the region of longitudinal curvature change (right). 
	The black and white solid lines identify the separation lines 
	in motion and the geometrical intersection line
	with the undisturbed free surface, respectively.}
   \label{fig:surface_pressure}
\end{figure}

In order to investigate the effects of the different
fuselage shapes on the hydrodynamics, the
$C_p$ contour maps over the wetted surface of {\bf S1}, {\bf S2} and {\bf S3} 
are compared in \autoref{fig:surface_pressure}. In particular,
the front part is shown on the
left side, whereas the rear part is shown on the right side.
For convenience, in the left-side plots the fuselages 
are aligned with respect to {\bf IG},
whereas in the right-side plots they are aligned 
with respect to \textbf{LZ} 
(see \autoref{fig:Ref_Frames} for their definitions).
%
The black and the white solid lines identify the separation line
and the geometrical intersection line with the undisturbed free surface
respectively. It is worth
reminding that {\bf S1} and {\bf S2} have a circular cross
section from the end of the nose
up to the point of longitudinal curvature change.
The lower part of the
cross section of {\bf S3} is elliptical and ``flatter'' than {\bf S2}.
The longitudinal curvature change at the rear is smooth in {\bf S1}, 
whereas it is quite abrupt for the shapes {\bf S2} and {\bf S3}.

As already observed in Section \ref{sec:jetshape}, the water impingement on 
the fuselage
body at the front leads to a high pressure region, which on its turn
induces a water pile-up and a spray, clearly displayed in all the fuselages. 
As expected, the pressure trends and values, as well as the separation
lines, for the fuselages {\bf S1} and
{\bf S2} are very similar in shape. 
The pressure in {\bf S2} appears slightly lower
than in {\bf S1}. 
This is due to the fact that the
spray root in \textbf{S2} lies at the end of the nose region, 
whereas in \textbf{S1} is in the region of constant longitudinal curvature,
see Figure \ref{fig:Fuselage_Shapes}. 
There are much more significant differences between 
the pressure trends of {\bf S1}/{\bf S2} and that of {\bf S3},
mainly as a results of the flatter shape,
and thus of the generally lower deadrise angle.
For this reason 
the pressure peak in the
shape {\bf S3} on the midline is higher than in
the shapes {\bf S1} and {\bf S2}, and so are the pressure values at the sides
in the spray root region. It is also observed that in {\bf S3} the distance 
between the spray
root and the spray tip (which is  approximately the distance from the location of 
maximum pressure to the separation line) is larger than in {\bf S1} and {\bf S2}
and is rather uniform moving from the midline
to the sides.

\begin{figure}[htbp]
	\centering
	\includegraphics[width=0.80\textwidth]{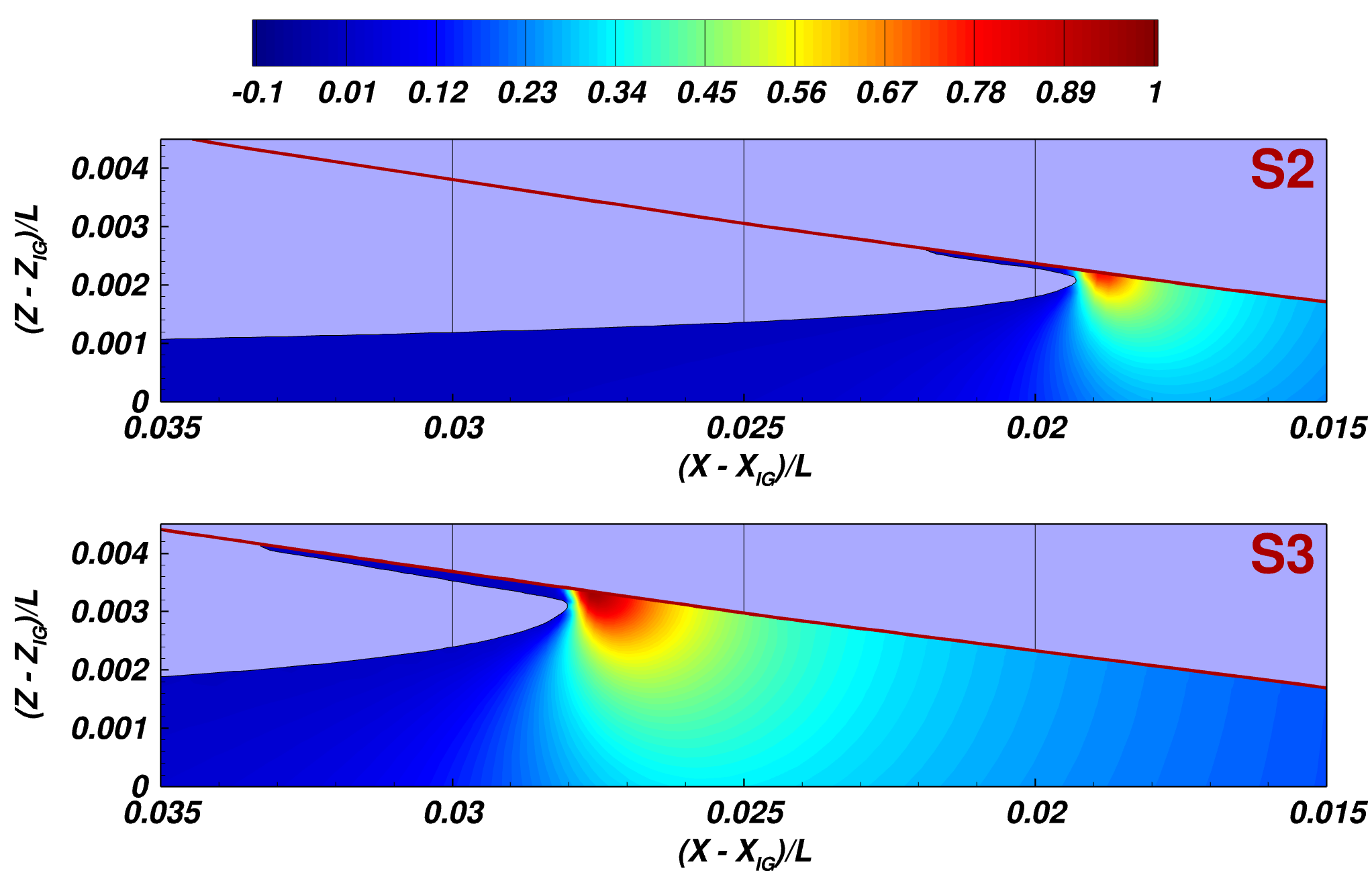}
	\caption{$C_p$ contours for the {\bf S2} and {\bf S3} fuselage shapes in the
  mid-plane delimited by the free surface. The red
	solid line represents the edge of the fuselage mid-plane section.}
	\label{fig:jetroot_zoom_S123_Cpcontours}
\end{figure}

In order to better visualise the effect of the transverse shape on the 
flow field, the $C_p$ contours and the free surface shapes
in the mid-plane (i.e. $Y/L=0$) are shown in \autoref{fig:jetroot_zoom_S123_Cpcontours}.
%
%
In these plots only the results for
{\bf S2} and {\bf S3} are shown, 
being the data for {\bf S1} very similar to those of {\bf S2}.
For convenience, the reference point for the
comparison is the $X$-coordinate of {\bf IG}.
In \autoref{fig:jetroot_zoom_S123_Cpcontours} 
the water pile-up in front of the geometrical
intersection point {\bf IG} is visible for both shapes,
with the thin water spray remaining well attached to the body surface.
As expected, the maximum pressure occurs approximately just behind the spray root,
with the pressure peak recorded for shape \textbf{S3} being 
higher than that for \textbf{S2}.
As a consequence, 
the spray extends slightly further ahead, and the water pile-up
is more relevant in {\bf S3}.
These effects are again related to the flatter lower surface
and to the reduced possibility of the fluid to escape from the sides.

\begin{figure}[htbp]
	\centering
  \includegraphics[width=0.85\textwidth]{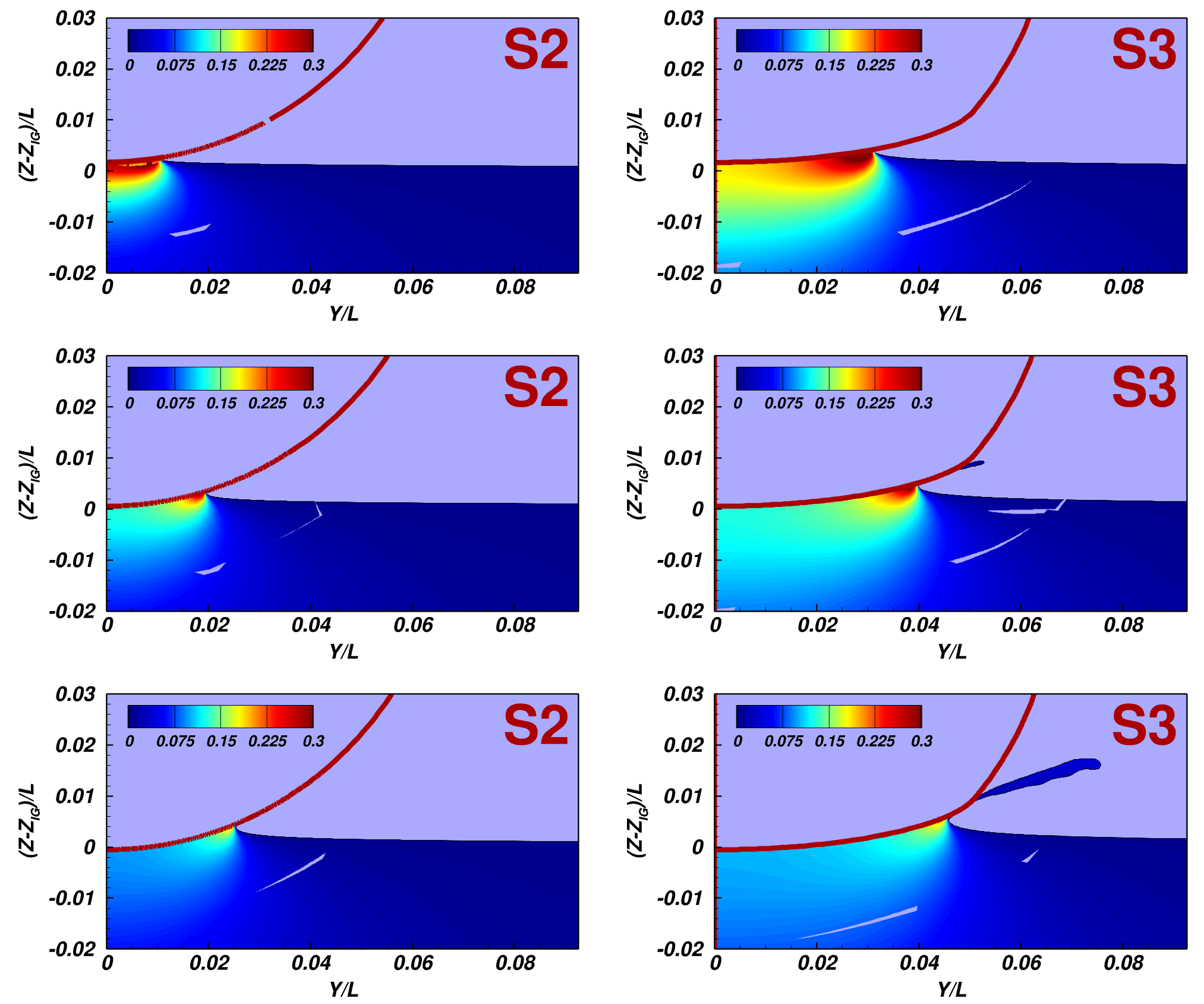}
	\caption{Free surface at different longitudinal sections for 
	the shapes \textbf{S2} and \textbf{S3}: $(X-X_{\textrm{IG}})/L=0.015$ (top)
	$(X-X_{\textrm{IG}})/L=0.005$ (middle) and $(X-X_{\textrm{IG}})/L=-0.005$ (bottom),
	coloured with the contour maps of the hydrodynamic pressure coefficient $C_p$. The red
	solid line represents the edge of the fuselage cross section at each
	specific $X$-location.}
	\label{fig:freesurface_slices_1}
\end{figure}

The evolution of the spray is also strongly affected by the
fuselage shape, as shown in \autoref{fig:freesurface_slices_1},
where  the $C_p$ contours and the free surface shape are plotted 
on several cross-sections along
the longitudinal direction.
%
%
At $(X-X_{\textrm{IG}})/L=0.015$,
i.e. ahead of \textbf{IG} the free surface is still
attached to the body for both shapes. 
As already anticipated, due to the smaller deadrise angle, 
the maximum pressure values for \textbf{S3} are higher then those for \textbf{S2}.
Moving backwards, i.e at $(X-X_{IG})/L=0.005$, the spray tip in
{\bf S3} reaches the point where the transverse shape profile changes 
from elliptical to circular, see \cite{iafrati2020experimental}.
The sudden change in the transverse curvature facilitates
the detachment of the spray from the body surface.
Further downstream, the water spray freely evolves under the effect of gravity.
Due to the more uniform curvature, in the {\bf S2} the separation of the 
water spray from the
wall takes place further downstream, as shown in
\autoref{fig:test_S2_slices_DX}.

The analysis of the flow field at the rear is more complex.
A comparison between the $C_p$ contour maps
in that region is shown in the right column of \autoref{fig:surface_pressure}.
It is evident that in all
cases, as a results of the change in the longitudinal curvature,
the flow accelerates ahead of 
\textbf{LZ}, hence the pressure decreases down to negative values, and
re-accelerates behind \textbf{LZ}, leading to a pressure increase. 
For the fuselage {\bf S1} the pressure reduction is very mild
and extends for a large distance behind {\bf LZ}.
On the contrary, in the fuselages {\bf S2} and {\bf S3}, as
a results of the abrupt curvature change, the flow acceleration
and deceleration ahead and behind \textbf{LZ} respectively
are much relevant, producing a significant and much more localised 
negative pressure area, in which the $C_p$ reaches values of
about -0.4. 
In {\bf S2} the area of negative pressure 
is smaller than in {\bf S3}, both ahead and behind \textbf{LZ},
and especially at the sides. This is clearly an effect of
the flatter bottom of the fuselage shape {\bf S3}.
\subsection{Forces acting on the fuselage}
\label{sec:forces}
\begin{figure}[htbp]
  \centering
\includegraphics[width=0.48\textwidth]{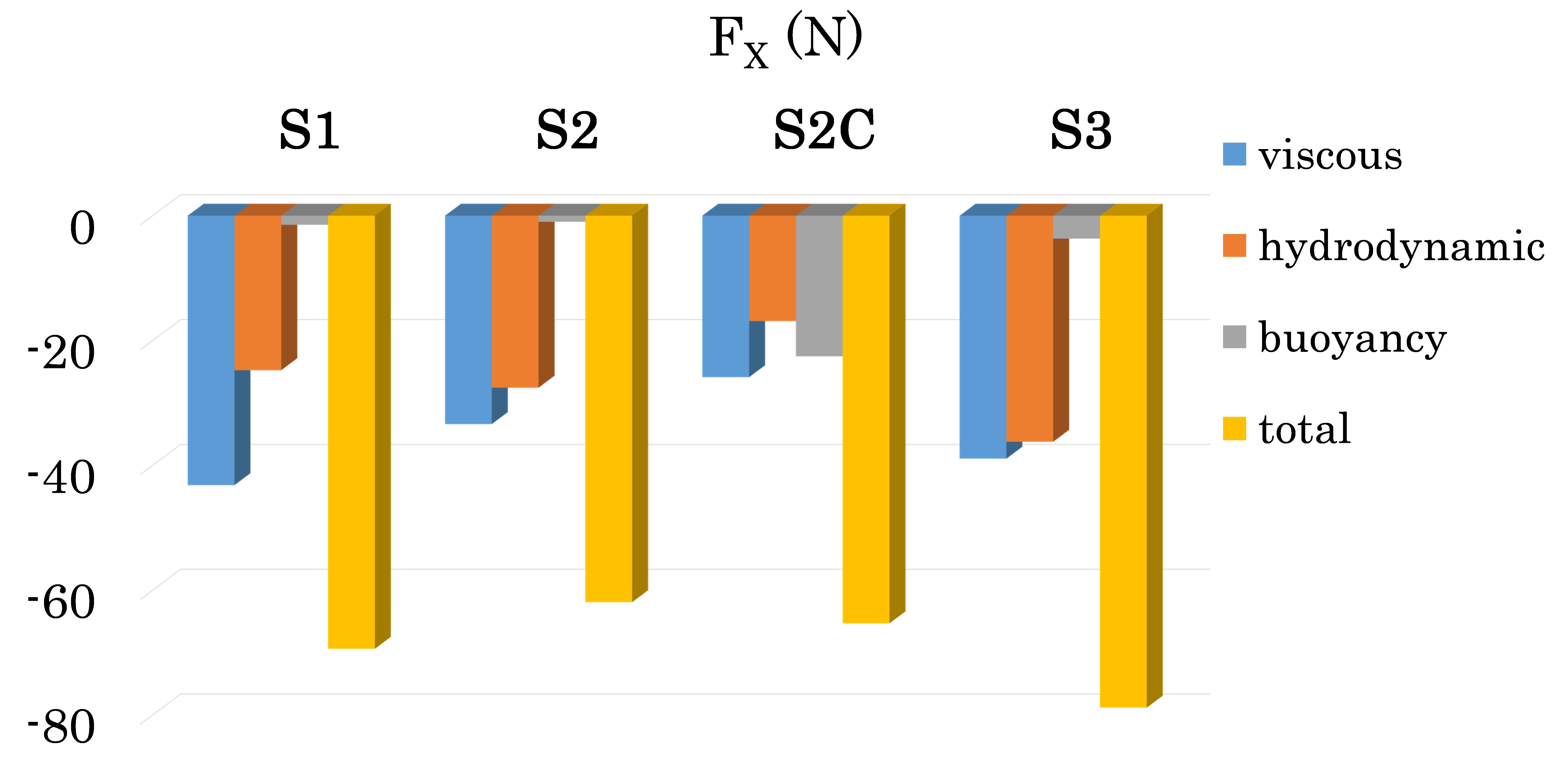}
\includegraphics[width=0.48\textwidth]{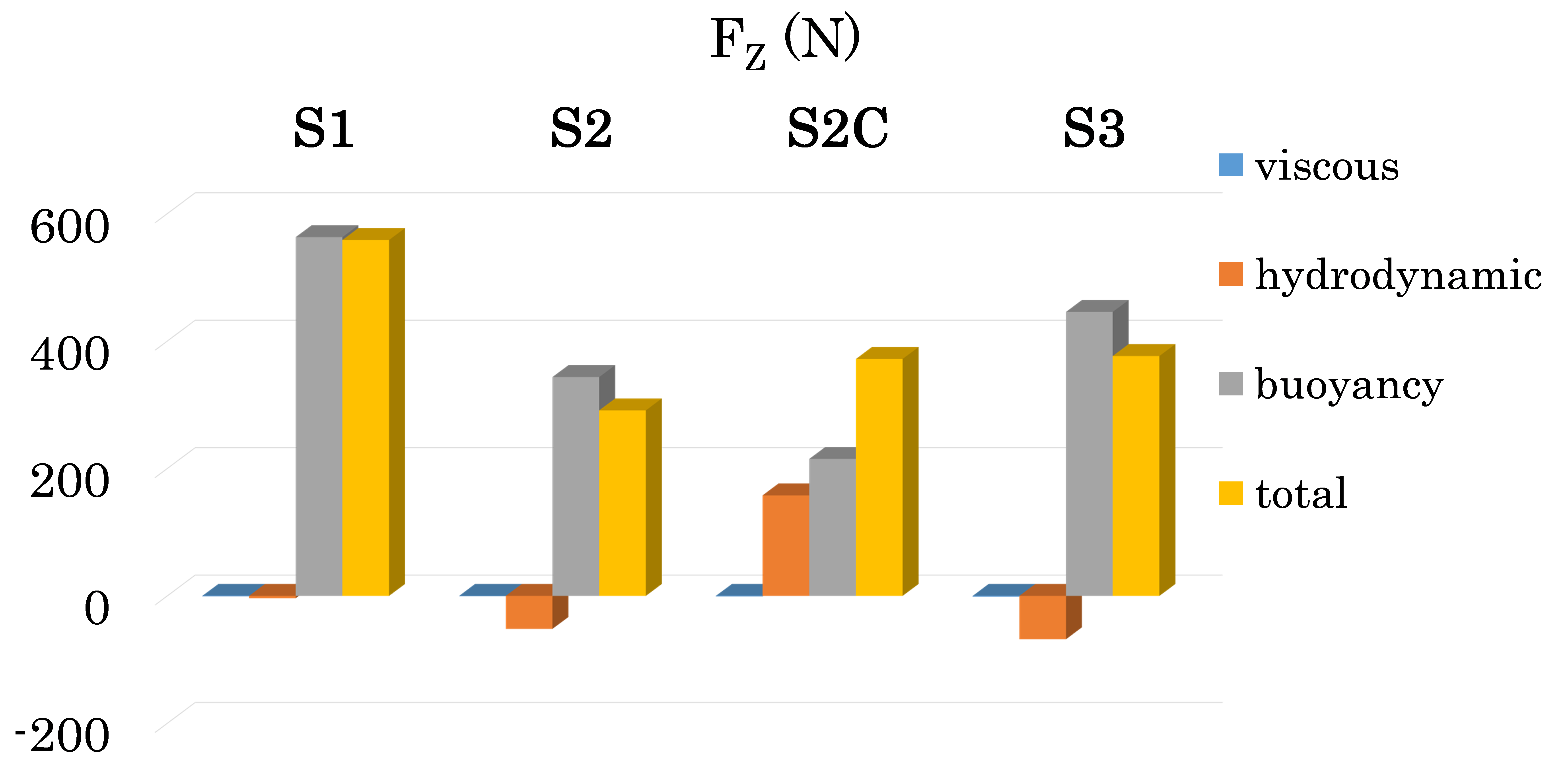}
 \caption{Histograms of the viscous, buoyancy and hydrodynamic contributions to $F_X$ (left) and $F_Z$ (right).}
  \label{fig:contributions}
\end{figure}

The forces acting on the fuselage
are derived by integration of the pressure and of the wall
shear stress fields over the wetted surface of the fuselages.
Three different contributions are distinguished,
namely the viscous contribution
(${\bf F_{v})}$,
the buoyancy contribution (${\bf F_{B}}$) and the pure hydrodynamic
pressure contribution (${\bf F_{D}}$), derived as:
\begin{equation}
  \begin{array}{l}
  \ds{{\bf F_{v}} = \rho U^2 L^2 \int_S \tau\,{\bf n} \,dS }   \\[3mm]
  \ds{{\bf F_{B}} = \rho U^2 L^2 \int_S \frac{z-z_{FS}}{Fr^2} \,{\bf n} \,dS }\\[3mm]
  \ds{{\bf F_{D}} = \rho U^2 L^2 \int_S p\,{\bf n} \,dS}
  \end{array}
	\label{eq:force_contributions}
\end{equation}
In \autoref{eq:force_contributions}, 
$\bf n$ is the normal (directed into the fluid) to the
wetted surface $S$, $p$ is the non-dimensional hydrodynamic pressure
and $\tau$ denotes the non-dimensional shear stress tensor.
The analysis is performed for the in-plane forces (i.e. the
projection along the $X$ and $Z$ axes of the different force contributions).
It is worth noting that in this case the force and the force
contributions are dimensional
quantities (in N).
The total force contributions for the four fuselages are reported in
\autoref{tab:force_contributions_num} and, graphically,
in the bar plots of \autoref{fig:contributions}.
\begin{table}[htb!]
	\centering
\caption{Total force contributions to $F_X$ and $F_Z$
for all the fuselage shapes. In the column ``viscous'' 
the flat plate analogy results are shown in round brackets.}
\label{tab:force_contributions_num}
\resizebox{0.75\textwidth}{!}{%
\begin{tabular}{|c|c|c|c|c|c|}
\hline
\multicolumn{1}{|l|}{} & \textbf{Shape} & \textbf{viscous} & \textbf{hydrodyn.} & \textbf{buoyancy} & \textbf{TOT} \\ \hline
\multirow{4}{*}{\textbf{$F_X$ [N]}} & {\bf S1} & -43.2 \emph{(-39.1)} & -24.8 & -1.5 & -69.4 \\ \cline{2-6} 
 & {\bf S2} & -33.4 \emph{(-30.0)} & -27.6 & -1.0 & -61.9 \\ \cline{2-6} 
 & {\bf S2C} & -25.9 \emph{(-22.4)} & -16.9 & -22.5 & -65.3 \\ \cline{2-6} 
 & {\bf S3} & -38.9 \emph{(-33.5)} & -36.2 & -3.7 & -78.8 \\ \hline\hline
\multirow{4}{*}{\textbf{$F_Z$ [N]}} & {\bf S1} & -0.8 & -3.7 & 563.1 & 558.6 \\ \cline{2-6} 
 & {\bf S2} & -0.5 & -51.8 & 343.5 & 291.2 \\ \cline{2-6} 
 & {\bf S2C} & -0.9 & 157.8 & 215.0 & 371.9 \\ \cline{2-6} 
 & {\bf S3} & -1.2 & -68.0 & 445.8 & 376.6 \\ \hline
\end{tabular}}
\end{table}
%
The viscous contribution in the $X$ direction for all the shapes 
is in a reasonably close agreement (although about 10\% lower)  
with the skin friction resistance $R_{FP}$ 
obtained using a flat plate analogy, in which the skin friction coefficient 
$C_{FP}$  is estimated as a function of Re$_L$ through the Hughes
correlation, 
see \cite{faltinsen2005hydrodynamics}, and
$R_{FP} = 1/2 \rho U^2 C_{FP} S_{wet}$.
The viscous and the hydrodynamic contribution to $F_X$
are of the same order of magnitude, whereas the buoyancy contribution
is negligible in all cases, as expected.
The ratio of the viscous contribution to 
the total value of $F_X$ spans
from about 50\% for the shapes {\bf S2} and {\bf S3} 
to 60\% for the shape {\bf S1}. This
is a consequence of the larger wetted surface
of \textbf{S1} compared to the other shapes.

As far as the $Z$-components are concerned, at this Froude number 
the hydrodynamic contribution 
is negative in all cases, except for the truncated fuselage shape {\bf S2C}. 
For the shape
\textbf{S1} the hydrodynamic contribution is very close to zero, whereas for shapes
\textbf{S2} and \textbf{S3} is about \mbox{-50~N and -70~N}, respectively. 
In all cases, except for the shape \textbf{S2C},
the hydrodynamic contribution is much lower in amplitude than 
the buoyancy contribution, which is positive and largely predominant. Therefore
the total force values result positive, i.e. oriented upwards. 
\subsubsection{Comparison with the experimental measurements}
The loads acting on the fuselage derived from the numerical simulations
are here validated against the measurements. A zeroing procedure
similar to that followed during the experiments has been performed numerically,
i.e. the hydrostatic loads, calculated at rest, are subtracted from 
those derived from the simulations of the fuselages in motion. 
The validation assessment 
is summarized in \autoref{tab:comparison_num_exp}.
\begin{table}[htbp]
\caption{Comparison between numerical and experimental loads.}
\label{tab:comparison_num_exp}
\resizebox{0.99\textwidth}{!}{%
\begin{tabular}{|c|c|c|c|c|c|}
\hline
\textbf{} & \textbf{Shape} & \textbf{Numerical} & \textbf{Experimental} & 
\textbf{Uncertainty} & \textbf{Deviation} \\ \hline
\multirow{4}{*}{\textbf{$F_X$ (N)}} & \textbf{S1} & -69.4 & -71.3 & 4.0\% & 2.69\% \\ \cline{2-6} 
 & \textbf{S2} & -61.7 & -68.2 & 4.0\% & 10.58\% \\ \cline{2-6} 
 & \textbf{S2C} & -65.2 &  &  &  \\ \cline{2-6} 
 & \textbf{\textbf{S3}} & -78.8 & -83.3 & 4.0\% & 5.73\% \\ \hline
 &  &  &  &  &  \\ \hline
\multirow{4}{*}{\textbf{$F_Z$ (N)}} & S1 & -78.0 & -48.2 & 22.0\% & 38.20\% \\ \cline{2-6} 
 & \textbf{S2} & -112.5 & -106.9 & 22.0\% & 4.99\% \\ \cline{2-6} 
 & \textbf{S2C} & 118.8 &  &  &  \\ \cline{2-6} 
 & \textbf{S3} & -123.1 & -128.2 & 22.0\% & 4.15\% \\ \hline
\end{tabular}
}
\end{table}

Having already compensated the water level lowering of 4~mm below 
the carriage during the run,
the experimental uncertainty reported in \autoref{tab:comparison_num_exp} 
takes into account the effect of a residual basin standing wave of amplitude 
3~mm.
The residual standing wave affects both the zeroing 
process and the time averages. To evaluate the first source of
uncertainty, the hydrostatic force acting on the \textbf{S2} fuselage at
different immersion depths, spanning a range of $\pm$ 6~mm about the
nominal one, are computed. The effect of a 
different water level during the run can be evaluated by computing the 
forces acting on the fuselage in motion at $U=5$~m/s at different
immersion depths, still spanning a range of $\pm$ 6~mm about the nominal one.
The force trends in the examined range are found to be linear.
Combining these analyses, an uncertainty of  
of 4 \% and 22 \% for the $X$ and the $Z$ force
components respectively are estimated.
These uncertainties are 
of the same order of magnitude of those estimated by analysing the acquired 
force time histories during the part of the carriage run
at constant speed.

The percentage deviations between the experimental and numerical results,
reported in the last columns of the \autoref{tab:comparison_num_exp}, 
are computed as:
\begin{equation*}
	\textrm{Deviation} = 100 \cdot \frac{F^{exp} - F^{num}}{F^{num}}
\end{equation*}
where $F^{exp}$ is the measured force value and
$F^{num}$ is the zeroed forces derived numerically.
The comparison denotes a fairly good agreement between the
CFD estimation and the experimental tests. The deviation for $F_X$ 
is of the same order of magnitude of the 
experimental uncertainty, except for the shape \textbf{S2},
around 10\%,
which can be considered nonetheless an acceptable value. 
Similarly, the deviation for $F_Z$ is much lower than the
corresponding experimental uncertainty, except for {\bf S1}, in which it 
is much larger than expected, being almost 40\%.
The reason for this discrepancy is due to the fact that
due to a technical issue, in this single test
the acquisition started just before the carriage departure, thus the
time interval to perform the zeroing at still carriage
results much lower than 1~s (the zeroing interval for all the other tests),
leading to a much higher uncertainty in the force readings.
\subsubsection{Sectional Force Distributions}
\label{sec:force_distributions}
\begin{figure}[htbp]
	\centering	
   \includegraphics[width=0.99\textwidth]{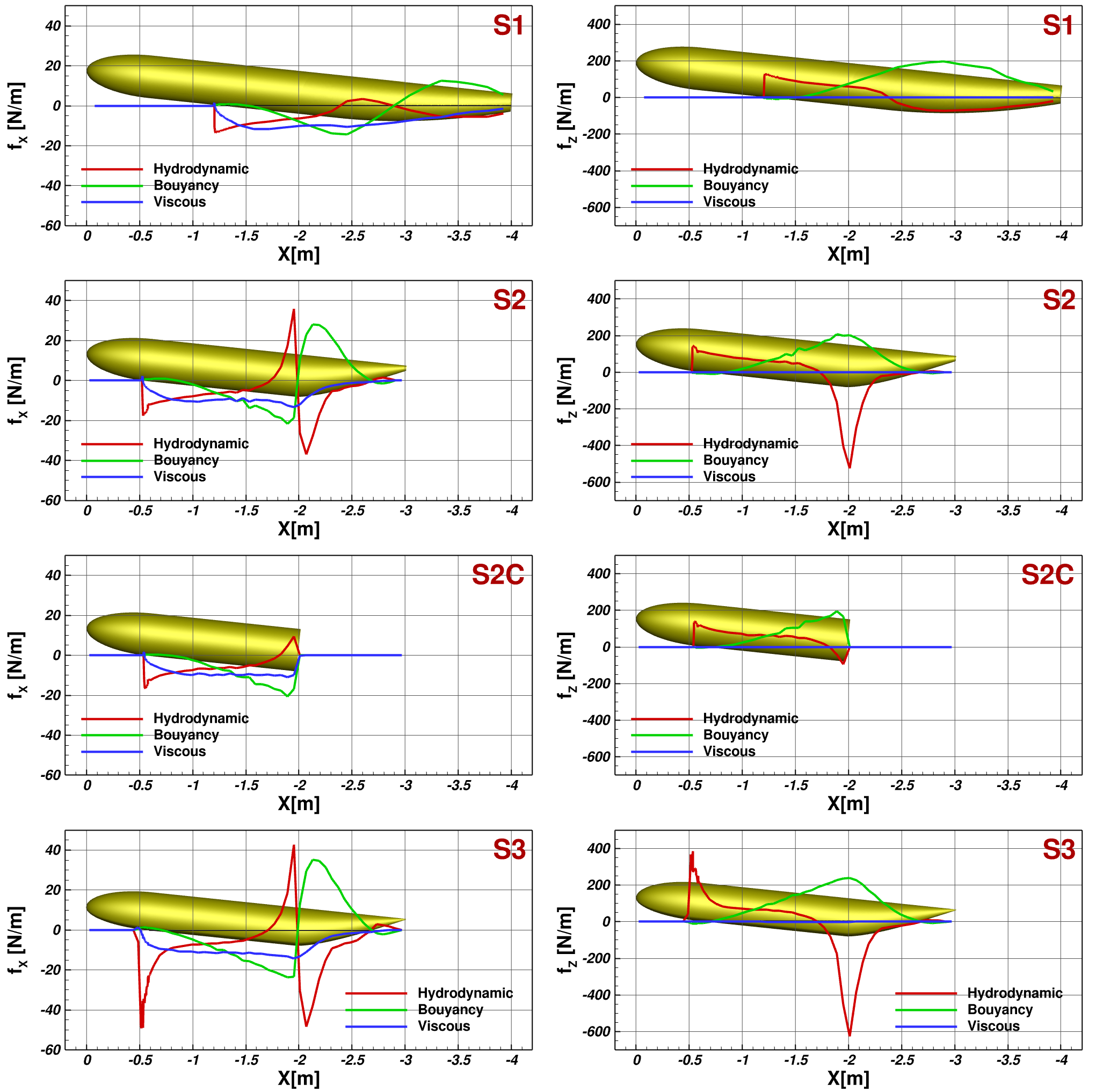}
	\caption{Hydrodynamic, buoyancy and viscous force distributions
	as a function of $X$, plotted with red, green and blue lines respectively,
	for all the fuselage shapes. The distributions relative to the
	$X$ and $Z$-forces are displayed in the left and right column, respectively.	
	The fuselage geometry is also drawn for the reader's convenience.}
	\label{fig:forze_distribuz_all}
\end{figure}

In order to explain the total values of the force components 
reported in \autoref{tab:force_contributions_num} and to gather
more information on the fuselage hydrodynamics, 
it is very informative to derive the 
sectional force distribution in the longitudinal direction per unit length,
denoted as $f_X(X)$ and $f_Z(X)$ for the four shapes. These distributions
are obtained by dividing the fuselage surface into several
strips in the longitudinal direction, computing the corresponding distribution 
of the force per each strip and dividing the value by the strip length.
The results are plotted in \autoref{fig:forze_distribuz_all}.
%
%

As expected, the viscous distribution $f_{X,V}$
(blue colour line, in the left column of \autoref{fig:forze_distribuz_all})
for the three shapes is always negative.
The strong pressure gradients at the curvature change for
the shapes {\bf S2} and {\bf S3} locally influence $f_{X,V}$
causing a slight decrease with a local minimum, 
which is however quite small compared to the base value. 
Such a result explains why the viscous contribution matches
the prediction of the flat plate analogy,
as confirmed by the values reported in \autoref{tab:force_contributions_num}.

The distribution $f_{Z,V}$
(blue colour line, in the right column of \autoref{fig:forze_distribuz_all})
is negligible everywhere
for all shapes, and the same happens for the integral value.

The overall buoyancy contribution in the $X$-direction
$F_{X,B}$ is very small as a result of the fact that 
the integral value of the buoyancy distribution $f_{X,B}$
(green colour line, in the left column of \autoref{fig:forze_distribuz_all})
for the  shapes {\bf S1}, {\bf S2} and {\bf S3}
ahead of $X_{\textrm{LZ}}$ is about equal but opposite in value to its
integral value behind $X_{\textrm{LZ}}$. 
On the contrary, for the truncated fuselage {\bf S2C}, in which the free-surface
abruptly separates at the stern, the negative overall integral value
of $f_{X,B}$ in the front part of {\bf S2C}, similar to that of {\bf S2},
is not counterbalanced by a positive contribution behind.

For the full fuselages
the values of $f_{Z,B}$
(green colour line, in the right column of \autoref{fig:forze_distribuz_all})
are all positive, with a maximum located
at about $X_{\textrm{LZ}}$, thus yielding a positive integral value.
For the fuselage {\bf S2C} the overall integral value $F_{Z,B}$
is lower than that of {\bf S2}, because after the maximum the
fuselage body is truncated.

The distribution $f_{Z,D}$
(red colour line, in the right column of \autoref{fig:forze_distribuz_all})
is related to the hydrodynamic
pressure variation shown in \autoref{fig:surface_pressure}, 
in which positive pressures appear at the front, about the spray root.
and negative pressures appear around the point where 
the longitudinal curvature changes.
For the shape {\bf S1} these two areas of opposite
sign are almost equivalent in the integral sense, leading to an almost null
hydrodynamic contribution $F_{Z,D}$, see again \autoref{tab:force_contributions_num}.
For the shapes {\bf S2} and {\bf S3},
instead, the negative integral value at the rear overcomes the positive
one at the spray root, leading to an overall negative hydrodynamic
contribution. In the shape {\bf S2C} the force distributions 
at the rear show very mild negative values
and for a very short extent, as a consequence of the pressure trend
observed in \autoref{fig:comparison_pressure_M2_M2cut},
thus yielding a generally positive total force in the $Z$-direction

As for $f_{X,D}$
(red colour line, in the left column of \autoref{fig:forze_distribuz_all}),
at the spray root the contribution of the 
positive pressure region to $F_{X,D}$
is negative, i.e. in the same sense of the drag force. At the
rear, looking at the distributions $f_{X,D}$ in \autoref{fig:forze_distribuz_all}
for the shapes \textbf{S1}, \textbf{S2} and \textbf{S3}, 
there is positive contribution ahead of the curvature change, followed by a
negative one behind it of about the same order of magnitude. Therefore, for all these shapes
the hydrodynamic $X$-contribution at the rear is almost
null, hence the overall values of $F_{X,D}$ (front + rear) are negative. 
Instead, for the shape {\bf S2C}, a positive contribution to $F_{X,D}$ ahead of the curvature change
is present, however its magnitude is lower than the
negative one at the spray root, resulting  in a negative overall $F_{X,D}$ value.
\section{Conclusions}
In this paper numerical simulations on the free surface flow around
three different fuselage models ({\bf S1}, {\bf S2} and {\bf S3}) moving in water at a
speed of 5~m/s, at a fixed pitch angle of 6$^{\circ}$ and immersion depth
146~mm are performed using the URANS level-set flow solver $\chi$navis.
The study is particularly focused on establishing
the role played by the fuselage shape on the hydrodynamics, 
pressure distributions and loads.

It is shown that a lower transverse curvature reduces 
the possibility for the fluid to escape from the sides 
and generates higher pressures and loads at the spray root.

Furthermore, the change in the longitudinal curvature occurring at
the rear causes an acceleration of the flow, and thus the generation 
of an extended region
of negative pressure. The presence of large suction forces 
in that region may explain
the increase in the pitch angle observed experimentally
in the free ditching tests.

As expected, the viscous contribution to the forces only affect the $X$-component,
and matches the estimates
based on the flat plate analogy.
The hydrodynamic and the buoyancy contributions, which derive from 
the integration of pressure over the wetted area, instead,
affect both the $X$ and $Z$ components.

A satisfactory agreement is found against the experimental data,
in spite of some uncertainties in the measurements due to the residual standing wave
and of the lowering of the water level causes by the moving carriage.

The simulations discussed in the present study are performed at low speed only.
Future activities will concern higher towing speeds, or higher Froude numbers,
in order to investigate the effects of speed on the hydrodynamics 
and the occurrence of phenomena like ventilation and/or cavitation,
which might substantially affect the aircraft dynamics at ditching.

\section*{Funding}
\label{sec:funding}
This project has been partly funded from the European Union's Horizon
2020 Research and Innovation Programme under Grant Agreement No. 724139
(H2020-SARAH: increased SAfety \& Robust certification for ditching of
Aircrafts \& Helicopters).

\section*{Acknowledgements}
\label{sec:acknowledgements}

The authors wish to thank Dr. Silvano Grizzi and Mr. Ivan Santic 
for their support during the experimental campaign.

%
\bibliographystyle{elsarticle-harv}
\bibliography{XNavis_paper_biblio}
\end{document}